\documentclass{IOS-Book-Article}

\usepackage{mathptmx}
\usepackage{amsmath,amssymb}
\usepackage{graphicx}
\usepackage{booktabs, multirow}
\usepackage{siunitx}
\usepackage{hyperref}
\usepackage{subcaption}
\usepackage{pdflscape}
\usepackage{threeparttable}
\usepackage[none]{hyphenat}
\usepackage{threeparttable}
\usepackage{pdflscape}   
\usepackage{hyperref}
\usepackage{adjustbox}   
\usepackage{array}
\numberwithin{equation}{section}
\numberwithin{equation}{section}
\usepackage[table]{xcolor}
\usepackage{colortbl}
\usepackage{booktabs}
\usepackage[table]{xcolor}
\usepackage{colortbl}
\usepackage{booktabs}
\usepackage{threeparttable}

\definecolor{StripeRow}{gray}{0.985}    
\definecolor{tablegold}{RGB}{245,242,230}   
\definecolor{headergold}{RGB}{235,230,210}  

\definecolor{TableHeader}{RGB}{232,241,250}
\definecolor{GroupRow}{RGB}{242,247,253}
\definecolor{StripeRow}{RGB}{250,252,255}
\definecolor{tblbg}{RGB}{248,246,238}

\definecolor{tblhead}{RGB}{236,232,216}

\definecolor{tblrule}{RGB}{170,165,150}
\arrayrulecolor{tblrule}

\hypersetup{
  colorlinks = true,
  linkcolor  = blue,
  citecolor  = blue,
  urlcolor   = blue
}

\begin{document}

\pagestyle{plain}

\begin{frontmatter}
\title{Timing of First Alzheimer's Neuroimaging: A Multicenter Analysis of Genetic, Cognitive, and Social Factors}

\author[A]{\fnms{Joshua} \snm{T. Korley}%
\thanks{Corresponding Author: Joshua Korley, Department of Epidemiology \& Biostatistics, University of South Carolina, Columbia, SC, USA; E-mail: jkorley@email.sc.edu}}
\and
\author[B]{\fnms{Enock} \snm{Adu Bonsu}}

\runningauthor{J. Korley and E. Adu Bonsu}

\address[A]{Department of Epidemiology \& Biostatistics, University of South Carolina, Columbia, SC, USA}
\address[B]{Department of Epidemiology \& Biostatistics, University of Arizona, Tucson, AZ, USA}

\begin{abstract}

\textbf{Background.}
Magnetic resonance imaging (MRI) is central to Alzheimer’s disease research and clinical evaluation, yet the timing of a participant’s first MRI after cohort entry varies substantially. In aging cohorts, many participants die before imaging, and treating death as noninformative censoring can distort inference on neuroimaging access and conceal systematic disparities.

\textbf{Objective.}
To quantify individual-level determinants and center-level heterogeneity in the timing of first neuroimaging, while accounting for the competing risk of death.

\textbf{Methods.}
We analyzed data from the National Alzheimer’s Coordinating Center Uniform Data Set v3 linked to MRI records from 2015 through March 2025. The study included 20,867 participants aged 50-95 years without dementia at baseline and with known APOE $\varepsilon4$ status. Time from baseline to first post-baseline MRI was modeled as the primary event, with death before MRI treated as a competing event. We fit Bayesian multilevel cause-specific accelerated failure time models with Alzheimer’s Disease Center random intercepts, using a Weibull model for MRI timing and a log-logistic model for death. Frequentist accelerated failure time and Cox models, including time-varying Cox models when assumptions failed, were used for sensitivity analyses.

\textbf{Results.}
APOE $\varepsilon4$ was not credibly associated with the timing of first MRI. Older age, lower educational attainment, and non-White race were associated with substantially delayed MRI acquisition. Center-level variation in MRI timing was substancial. Conversely, APOE $\varepsilon4$ homozygosity was associated with earlier death prior to MRI, alongside strong effects of age and cognitive impairment.

\textbf{Conclusions.}
The timing of first MRI in Alzheimer’s research cohorts is driven primarily by demographic, social, and center-level factors rather than genetic risk. Accounting for competing mortality and site heterogeneity is necessary to avoid biased neuroimaging samples and to support equitable Alzheimer’s research.
\end{abstract}

\begin{keyword}
Alzheimer's disease\sep MRI timing\sep competing risks\sep Bayesian multilevel model\sep accelerated failure time model\sep health disparities\sep site heterogeneity
\end{keyword}

\end{frontmatter}

\thispagestyle{empty}
\pagestyle{empty}

\clearpage 
\pagestyle{plain}
\setcounter{page}{1}
\section{Introduction}

Neuroimaging plays a pivotal role in Alzheimer's disease (AD) clinical evaluation and research. Structural magnetic resonance imaging (MRI) supports etiologic assessment, characterization of neurodegeneration, and exclusion of alternative pathologies, and it is routinely used to define cohorts and derive imaging biomarkers in observational studies and clinical trials \cite{Jack2010,Frisoni2010,Weiner2013,Jack2018}. In many workflows, MRI is also an essential requirement: participants without neuroimaging are often excluded from downstream analyses of brain structure, biomarker trajectories, and treatment response. For this reason, the timing of first neuroimaging is not merely administrative. It is a selection mechanism that affects representativeness of imaging enriched samples and can influence the external validity of imaging based inferences.

Despite its importance, the process by which participants receive their first brain MRI after cohort entry is not well characterized in multicenter AD research settings. Imaging is sometimes assumed to occur early and with considerable uniformity, driven primarily by biological risk and cognitive status. In practice, access to neuroimaging reflects a complex interaction among patient characteristics, clinical decision making, referral pathways, and institutional capacity. Structural barriers and differential access to specialty care can translate into delays in advanced diagnostic procedures, including neuroimaging, and such delays can contribute to inequitable participation in research cohorts that rely on imaging derived endpoints \cite{Saadi2017,Vagal2023,Williams1997,Braveman2011}.
Recent systematic reviews have quantified substantial underrepresentation of racial and ethnic minorities in AD neuroimaging studies, with median representation of 88.9\% White participants and only 7.3\% Black/African American participants \cite{Lim2023}. NACC participants themselves differ systematically from nationally representative samples in sociodemographic and health characteristics, with differences amplified across racial and ethnic groups \cite{Arce2023}. Moreover, prediction models in AD research must explicitly account for the competing risk of death, as many risk factors for AD are also risk factors for mortality, and some high-risk patients may die before developing symptomatic disease \cite{Barnes2012}.

A fundamental complication in studying time to first MRI is that MRI is not guaranteed to occur. In older cohorts with substantial comorbidity, a nontrivial fraction of participants die before receiving neuroimaging. Once death occurs, a participant can never receive MRI. Death therefore acts as a terminal competing event rather than a benign loss to follow up. Treating death as simple right censoring, as in naive survival analyses, requires an implausible independence assumption and can distort estimates of imaging incidence and covariate effects \cite{Pintilie2006,Beyersmann2012,KalbfleischPrentice2011}. Competing risk methods provide a coherent framework for separating the probability of receiving MRI from the probability of dying before MRI and for estimating covariate effects on each process.

Most competing risk analyses in AD have relied on proportional hazards models, including cause specific Cox regression and subdistribution hazard models \cite{Fine1999,Pintilie2006,Beyersmann2012,KalbfleischPrentice2011,Barnes2012}. While widely used, these approaches can be difficult to interpret when proportional hazards assumptions fail. In heterogeneous clinical cohorts, effects of age, education, and race on healthcare utilization and survival can evolve over follow up time, and nonproportionality is common \cite{GrambschTherneau1994,Stensrud2020}. When hazards are not proportional, a single hazard ratio can obscure clinically relevant dynamics and complicate interpretation in access to care settings where delays are central.

Accelerated failure time (AFT) models provide a complementary and often more policy relevant perspective for studying diagnostic timing. Rather than modeling instantaneous risk, AFT models describe how covariates stretch or compress the time scale of an event, yielding time ratios that directly quantify delay or acceleration \cite{Wei1990,KalbfleischPrentice2011}. For neuroimaging, this interpretation maps to the scientific question: how much longer do certain groups wait to receive their first MRI, and how quickly does death preclude imaging in high risk subgroups. Thus, AFT can be adopted in AD health services questions, particularly when cause-specific competing risks, such as death, influence the observation of the primary event of interest like receiving an MRI.

Another key and often overlooked feature of multicenter cohorts is the potential for strong site level heterogeneity. The National Alzheimer's Coordinating Center (NACC) aggregates data from many Alzheimer's Disease Centers (ADCs) operating in distinct health systems with differing referral environments, imaging capacity, and operational workflows \cite{Morris2006,Beekly2007}. Recent work has documented substantial between-center heterogeneity in recruitment characteristics and participant profiles across ADCs, raising concerns about external validity and generalizability of findings from NACC data \cite{Chan2025,Arce2023}. If site differences are large, pooled models that ignore clustering can obscure individual level associations with structural differences across centers. This can misrepresent disparities and reduce transportability, particularly as imaging availability varies by center.

Bayesian hierarchical survival models provide a principled approach to address competing risks and site heterogeneity simultaneously. Multilevel survival models extend standard survival analysis by incorporating cluster-specific random effects to account for hierarchical data structures common in health services research \cite{Austin2017,Yang2009}. Random effects for ADCs can separate within center associations from between center differences and provide direct quantification of site level variability in neuroimaging timing. Bayesian estimation also yields coherent uncertainty quantification for heavy tailed distributions that may better capture mortality related event times in older populations \cite{Gelman2013,Carpenter2017,Buerkner2017}.

Within this context, we use NACC UDS version-3 linked MRI records to study the timing of first brain MRI among participants without dementia at baseline. We treat death prior to MRI as a competing event and model time to first MRI and time to death before MRI using cause specific AFT models. The primary analysis employs Bayesian multilevel AFT models with ADC specific random intercepts to address two aims with equal emphasis: to quantify individual level determinants of neuroimaging timing and to quantify center level heterogeneity in the timing of first neuroimaging. We additionally report competing risk summaries via cumulative incidence functions and conduct sensitivity analyses using frequentist AFT models and Cox models with time varying effects to assess robustness.

\section{Methods}

\subsection{Data source and study population}

We analyzed participants from the NACC Uniform Data Set (UDS) version-3 linked with MRI records, spanning 2015 through March 2025. NACC harmonizes standardized demographic, clinical, and cognitive assessments collected at ADCs across the United States using common protocols \cite{Morris2006,Beekly2007}. Participants contribute repeated visits, typically annually, with longitudinal recording of cognitive testing, clinical status, and selected health history.

Baseline was defined as each participant's first eligible UDS v3 visit during the study window. We included participants aged 50 to 95 years at baseline who were free of dementia at baseline and had known APOE $\varepsilon4$ genotype coded as 0, 1, or 2 copies. We excluded observations with nonpositive follow up time or inconsistent timing information that prevented construction of a valid time to event outcome. Participants were followed from baseline until the earliest of first MRI, death prior to MRI, or last recorded contact. Because participants are clustered within ADCs, the analytic dataset retained the center identifier (NACCADC) to support hierarchical modeling of site level differences.

\subsection{Ascertainment of timing of first neuroimaging}

The primary endpoint was time from the baseline visit date to the first observed post baseline brain MRI. Timing was extracted by linking UDS records to the MRI linked file and identifying, for each participant, the earliest MRI date occurring on or after baseline. Time to MRI was computed in years as the difference between the first post baseline MRI date and the baseline visit date divided by 365.25. For participants with multiple MRI records, only the earliest post baseline MRI was used. Participants with MRI dates prior to baseline were not counted as having the event at time zero because the estimand targets acquisition of the first MRI after cohort entry into the eligible non demented baseline state. Participants with no post baseline MRI were considered at risk until death or censoring.

\subsection{Outcome definition and competing risk structure}

Let $T_i$ denote observed follow up time from baseline for participant $i$ in years. Let $J_i$ denote event type, with $J_i=1$ for receipt of first post baseline brain MRI, $J_i=2$ for death occurring before any post baseline MRI, and $J_i=0$ for right censoring at last known follow up in the absence of MRI and death. Death prior to MRI permanently precludes MRI, so it is a terminal competing event rather than a censoring mechanism \cite{Pintilie2006,Beyersmann2012,KalbfleischPrentice2011}. Our primary inferential strategy therefore models the time to each cause in a cause specific framework, while descriptive summaries use cumulative incidence functions (CIFs) to partition the probability of MRI and death over time in a manner coherent under competing risks.

\subsection{Exposure and baseline covariates}

The main exposure was APOE $\varepsilon4$ allele count, modeled categorically with three levels: 0 (reference), 1, and 2 copies. APOE $\varepsilon4$ is the most common genetic risk factor for late onset AD and is associated with earlier disease onset and progression, motivating assessment of whether genetic risk influences neuroimaging timing in real world ADC settings \cite{Liu2013}.

Baseline covariates were selected to capture demographic, social, and clinical factors plausibly related to neuroimaging access and mortality. Age at baseline was categorized as under 65, 65 to 79, and at least 80 years. Education was categorized as high school or less (reference), undergraduate (up to 16 years), and graduate (more than 16 years) as a pragmatic proxy for socioeconomic resources and health system navigation. Sex was included as female versus male. Race was modeled as White (reference), Black, and Other, consistent with available coding. Baseline cognitive and functional status was represented by the Clinical Dementia Rating (CDR) global score dichotomized as 0 versus at least 1 to capture impairment among participants without baseline dementia. Living situation at baseline was modeled as living alone (reference), living with spouse or partner, and living in another household arrangement. Vitamin B12 deficiency status was included as a baseline clinical history marker potentially related to frailty and clinical evaluation intensity.

All covariates were measured at baseline to preserve a clear temporal ordering between predictors and subsequent imaging or death, and to align the estimand with baseline risk profiling rather than time updated decision making.

\subsection{Descriptive analyses and competing risk summaries}

Baseline characteristics were summarized overall and by event type (MRI, death before MRI, censored). Continuous variables were reported as mean and standard deviation and compared using Welch two sample tests. Categorical variables were summarized as counts and percentages and compared using Pearson chi square tests, with Fisher exact tests used when expected counts were sparse.

To describe the competing risk process, we reported event counts and plotted cumulative incidence functions for first MRI and death before MRI. CIFs provide coherent estimates of the probability of each event over time in the presence of competing risks, unlike Kaplan Meier estimators that treat competing events as censoring \cite{KalbfleischPrentice2011,Beyersmann2012}.
\subsection{Statistical Analysis}
\subsubsection*{Primary modeling strategy: cause-specific accelerated failure time models}

The primary inferential framework consisted of cause-specific accelerated failure time (AFT) models for each cause $j \in \{1,2\}$. This approach was selected because the scientific quantities of interest are delays or accelerations in the timing of first neuroimaging and death, which are naturally expressed on the time scale. In AFT models, regression coefficients admit direct interpretation through time ratios, providing a transparent measure of temporal acceleration or delay \cite{Wei1990}.

Cause-specific models were fitted separately for each event type. For a given cause, individuals experiencing the competing event were treated as censored at their observed time, consistent with standard cause-specific modeling in the presence of competing risks \cite{Pintilie2006,KalbfleischPrentice2011}. This strategy allows covariate effects on each event process to be examined independently while respecting the irreversible nature of death prior to imaging. Throughout, the term “cause-specific” refers to separate modeling of the time scale for each event type, rather than to subdistribution or cumulative incidence regression.

\subsubsection*{Frequentist AFT formulation}

For cause $j$, the AFT model was specified as
\begin{equation}
\log(T_i) = \eta_{ij} + \sigma_j \epsilon_{ij},
\qquad
\eta_{ij} = \beta_{0j} + \mathbf{x}_i^\top \boldsymbol{\beta}_j,
\label{eq:aft-frequentist}
\end{equation}

where $\mathbf{x}_i$ denotes the vector of baseline covariates for participant $i$, $\sigma_j > 0$ is a scale parameter, and $\epsilon_{ij}$ is an error term whose distribution determines the parametric family on the original time scale. Covariate effects are summarized using time ratios. For covariate $k$, the time ratio is defined as
\begin{equation}
\mathrm{TR}_{jk} = \exp(\beta_{jk}),
\label{eq:time-ratio}
\end{equation}

where $\mathrm{TR}_{jk} > 1$ indicates a longer expected time to the event (delay) and $\mathrm{TR}_{jk} < 1$ indicates an earlier occurrence (acceleration), relative to the reference category with other covariates held fixed.

\subsubsection*{Bayesian multilevel AFT formulation with site-level random effects}

The primary analysis extended the AFT framework to a Bayesian multilevel setting to account for heterogeneity across Alzheimer’s Disease Centers (ADCs). Multilevel survival models explicitly disentangle total variation into components at different levels of the data hierarchy, enabling identification of sources of variation for targeted interventions \cite{Austin2017,Yang2009}. For participant $i$ enrolled at center $c(i)$ and cause $j$, the model was specified as
\begin{equation}
\log(T_i) = \alpha_{0j} + u_{c(i),j} + \mathbf{x}_i^\top \boldsymbol{\beta}_j + \sigma_j \epsilon_{ij},
\label{eq:aft-bayesian}
\end{equation}

where $u_{c(i),j}$ represents a center-specific random intercept capturing systematic differences in baseline timing across ADCs after adjustment for measured covariates. These random effects were modeled as
\begin{equation}
u_{c,j} \sim \mathcal{N}(0,\tau_j^2),
\label{eq:center-re}
\end{equation}

with $\tau_j$ quantifying between-center heterogeneity on the log-time scale. Larger values of $\tau_j$ indicate greater structural variation across centers in the timing of first MRI or death prior to MRI that cannot be explained by individual-level characteristics alone.

Distributional assumptions were specified through the error term. For Cause 1, a Weibull AFT model was used, corresponding to an extreme value distribution for $\epsilon_{i1}$ and yielding a Weibull distribution for event times on the original scale. This choice accommodates monotone hazard behavior consistent with referral and scheduling processes underlying MRI acquisition. For Cause 2, a log-logistic AFT model was employed, corresponding to a logistic distribution for $\epsilon_{i2}$ and yielding a log-logistic distribution for event times, which allows for heavier tails and nonmonotone hazards commonly observed in mortality data from aging cohorts \cite{KalbfleischPrentice2011,Pintilie2006}.

\subsubsection*{Priors and computation}

Weakly informative priors were used to stabilize estimation while allowing the data to dominate inference in this large cohort. Regression coefficients were assigned symmetric, mean-zero normal priors to reflect the belief that extreme effects on the log-time scale are unlikely a priori. Intercepts were given broader normal priors to accommodate variation in baseline timing across causes. All strictly positive parameters, including scale parameters and the standard deviations of center-level random effects, were assigned priors with support restricted to the positive real line.

To evaluate sensitivity to prior specification, the primary Bayesian models were refitted under a structured grid of alternative prior sets that varied the degree of shrinkage on regression coefficients and the tail behavior of priors on variance and distributional parameters. These analyses were designed to assess robustness rather than to explore alternative inferential targets. Full prior definitions and comparative results are reported in the Supplementary Material (Table~\ref{tab:prior-sens-tr}).
Models were estimated in R using \texttt{brms} with the \texttt{cmdstanr} backend. Four Markov chains were run with 4{,}000 iterations per chain, including 2{,}000 warmup iterations. Convergence and sampling efficiency were evaluated using split-chain $\widehat{R}$ statistics, effective sample sizes, trace plots, autocorrelation diagnostics, and posterior density overlays.

\subsubsection{Model selection and justification of distributional choices}

Candidate parametric AFT distributions were evaluated separately for each cause. Considered families included Weibull, log-normal, and log-logistic models. Selection was informed by information criteria from preliminary frequentist analyses and by graphical assessment of survival and residual-based diagnostics, with emphasis on stability and interpretability for each cause-specific process (Figures~\ref{fig:aft-mri}-\ref{fig:aft-death}).

The final Bayesian models employed a Weibull AFT specification for time to first MRI and a log-logistic AFT specification for time to death prior to MRI. These choices provided the best empirical fit for each outcome and yielded consistent inference across frequentist and Bayesian frameworks. Using distinct distributions for the two causes reflects differences in underlying mechanisms: imaging timing is shaped by referral and scheduling pathways that tend to induce monotone hazard behavior, whereas mortality processes in aging cohorts often exhibit heavier tails and greater heterogeneity.

\subsubsection{Sensitivity analysis}
\label{sec:sensitivity}

We assessed robustness to both modeling choices and prior specification.
First, we fit frequentist cause-specific AFT models as a likelihood-based benchmark and cause-specific Cox models as a hazard-scale comparator. When Schoenfeld residuals indicated nonproportional hazards, we fit time-varying coefficient Cox models using log-time interactions to describe how relative hazards evolved over follow-up.

Second, we conducted a structured prior sensitivity analysis for the Bayesian multilevel AFT models. The outcome definitions, covariates, and reference categories were held fixed, while priors on regression coefficients, intercepts, ADC-level random-effect standard deviations, and distributional parameters (Weibull shape; log-logistic scale) were varied across a prespecified grid (Table~\ref{tab:prior-sens-tr}). Across all prior sets, posterior time ratios and 95\% credible intervals were materially unchanged, with no covariate exhibiting a reversal in direction or a practically meaningful shift in magnitude. These results indicate that the primary inferences are driven by the observed data rather than by a single prior specification.\\

All statistical analyses were performed in RStudio
(R version 4.4.2), using the survival package for competing risk models, ggplot2 and survminer for forest plots. Bayesian models were fit using \texttt{brms} with the \texttt{cmdstanr} backend. Survival utilities used \texttt{survival} \cite{Therneau2023}.

\section{Results}

\subsection{Cohort characteristics and event distribution}

\begin{figure}[!ht]
\centering
\includegraphics[width=0.70\textwidth]{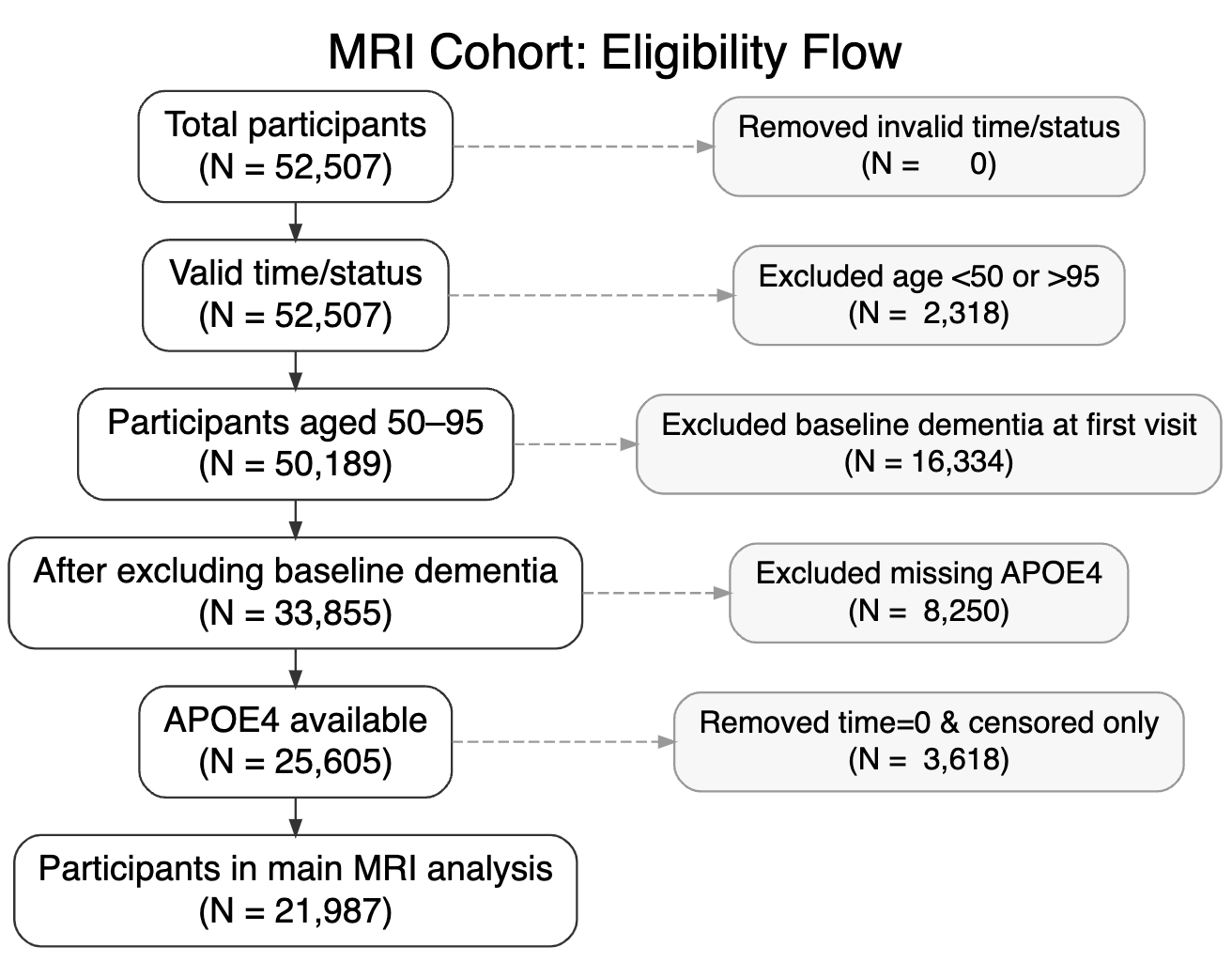}
\caption{Eligibility flow for the MRI analytic cohort.}
\label{fig:eligibility}
\end{figure}

The final analytic cohort consisted of 20{,}867 participants without dementia at baseline who met all inclusion criteria and contributed valid follow-up for neuroimaging and mortality outcomes (Figure~\ref{fig:eligibility}). Baseline characteristics stratified by 
observed event type are summarized in  Table~\ref{tab:baseline-event}. Overall, the cohort had a mean age of 72.2 years (SD 9.8) and a mean educational attainment of 15.5 years (SD 3.1). Cognitive and functional status at baseline reflected a predominantly mildly impaired population, with a mean CDR-SUM of 3.1 (SD 4.4).

\begin{table}[!ht]
\centering
\caption{Baseline characteristics by event type: first MRI, death before MRI, and overall cohort.}
\label{tab:baseline-event}
\scriptsize
\begin{threeparttable}

\setlength{\tabcolsep}{6pt}
\renewcommand{\arraystretch}{1.15}

\begin{tabular}{
>{\columncolor{tblbg}}l
>{\columncolor{tblbg}}c
>{\columncolor{tblbg}}c
>{\columncolor{tblbg}}c
>{\columncolor{tblbg}}c
}
\toprule
\rowcolor{tblhead}
\textbf{Characteristic} &
\textbf{P-value} &
\textbf{First MRI} &
\textbf{Death before MRI} &
\textbf{Overall} \\
\midrule

\textbf{Age (years), mean (SD)} &
$<0.001$ &
69.4 (8.7) &
74.6 (10.1) &
72.2 (9.8) \\
Missing (\%) & & 0 (0.0\%) & 0 (0.0\%) & 0 (0.0\%) \\[2pt]

\textbf{Education (years), mean (SD)} &
$<0.001$ &
15.9 (3.0) &
15.2 (3.2) &
15.5 (3.1) \\
Missing (\%) & & 0 (0.0\%) & 0 (0.0\%) & 0 (0.0\%) \\[2pt]

\textbf{CDR-SUM, mean (SD)} &
$<0.001$ &
1.2 (2.2) &
4.7 (5.1) &
3.1 (4.4) \\
Missing & & 0 (0.0\%) & 0 (0.0\%) & 0 (0.0\%) \\[4pt]

\textbf{APOE $\varepsilon4$} & $<0.001$ & & & \\
\hspace{0.4cm}No $\varepsilon4$ & &
6349 (60.5\%) & 6869 (55.8\%) & 13218 (58.0\%) \\
\hspace{0.4cm}One $\varepsilon4$ & &
3478 (33.2\%) & 4448 (36.1\%) & 7926 (34.8\%) \\
\hspace{0.4cm}Two $\varepsilon4$ & &
662 (6.3\%) & 999 (8.1\%) & 1661 (7.3\%) \\
Missing & & 0 (0.0\%) & 0 (0.0\%) & 0 (0.0\%) \\[4pt]

\textbf{Sex} & $<0.001$ & & & \\
\hspace{0.4cm}Male & &
4226 (40.3\%) & 6319 (51.3\%) & 10545 (46.2\%) \\
\hspace{0.4cm}Female & &
6263 (59.7\%) & 5997 (48.7\%) & 12260 (53.8\%) \\
Missing & & 0 (0.0\%) & 0 (0.0\%) & 0 (0.0\%) \\[4pt]

\textbf{Race/Ethnicity} & $<0.001$ & & & \\
\hspace{0.4cm}Hispanic & &
1004 (9.6\%) & 501 (4.1\%) & 1505 (6.6\%) \\
\hspace{0.4cm}NH Asian/Other & &
324 (3.1\%) & 194 (1.6\%) & 518 (2.3\%) \\
\hspace{0.4cm}NH Black & &
1639 (15.6\%) & 1150 (9.3\%) & 2789 (12.2\%) \\
\hspace{0.4cm}NH White & &
7522 (71.7\%) & 10471 (85.0\%) & 17993 (78.9\%) \\
Missing & & 0 (0.0\%) & 0 (0.0\%) & 0 (0.0\%) \\[4pt]

\textbf{B12 deficiency} & 0.102 & & & \\
\hspace{0.4cm}Absent &
& 9762 (93.1\%) & 11530 (93.6\%) & 21292 (93.4\%) \\
\hspace{0.4cm}Recent/Active &
& 567 (5.4\%) & 590 (4.8\%) & 1157 (5.1\%) \\
\hspace{0.4cm}Remote/Inactive &
& 160 (1.5\%) & 196 (1.6\%) & 356 (1.6\%) \\
Missing & & 0 (0.0\%) & 0 (0.0\%) & 0 (0.0\%) \\[4pt]

\textbf{Living situation} & $<0.001$ & & & \\
\hspace{0.4cm}Alone &
& 1585 (25.8\%) & 1233 (29.9\%) & 2863 (23.5\%) \\
\hspace{0.4cm}Spouse/Partner &
& 3774 (61.3\%) & 2300 (55.8\%) & 6800 (55.7\%) \\
\hspace{0.4cm}Other household &
& 632 (10.3\%) & 275 (6.7\%) & 930 (7.6\%) \\

\bottomrule
\end{tabular}

\begin{tablenotes}[flushleft]\scriptsize
\item SD denotes standard deviation. Event type is defined by first observed post-baseline MRI (Cause 1), death prior to any post-baseline MRI (Cause 2), or censoring at last contact.
\end{tablenotes}

\end{threeparttable}
\end{table}

Participants who died before receiving a post-baseline MRI differed systematically from those who underwent imaging. This group was older at baseline (mean 74.6 vs 69.4 years), and greater functional impairment (mean CDR-SUM 4.7 vs 1.2), indicating a clinically more vulnerable subgroup with elevated competing mortality. In contrast, participants who received MRI had higher educational attainment (mean 15.9 vs 15.2 years) and were more likely to be living with a spouse or partner.

The cohort was 46.2\% male and 53.8\% female overall. Racial and ethnic composition reflected the demographics of ADC enrollment, with 78.9\% White participants, 12.2\% Black participants, and smaller proportions of Hispanic and other racial groups. APOE $\varepsilon4$ carriage was common, with 38.0\% of participants carrying at least one $\varepsilon4$ allele and 7.3\% homozygous. These characteristics highlight substantial heterogeneity in demographic, social, and clinical factors that may influence both neuroimaging access and competing mortality.

During follow-up, a substantial proportion of participants either experienced death prior to MRI or were censored without imaging, emphasizing that first post-baseline MRI was not guaranteed even within ADC follow-up systems. This event structure underscores the importance of explicitly accounting for competing mortality when analyzing the timing of neuroimaging.

\subsection{Competing risk summaries}

\begin{figure}[!ht]
\centering
\includegraphics[width=0.7\textwidth]{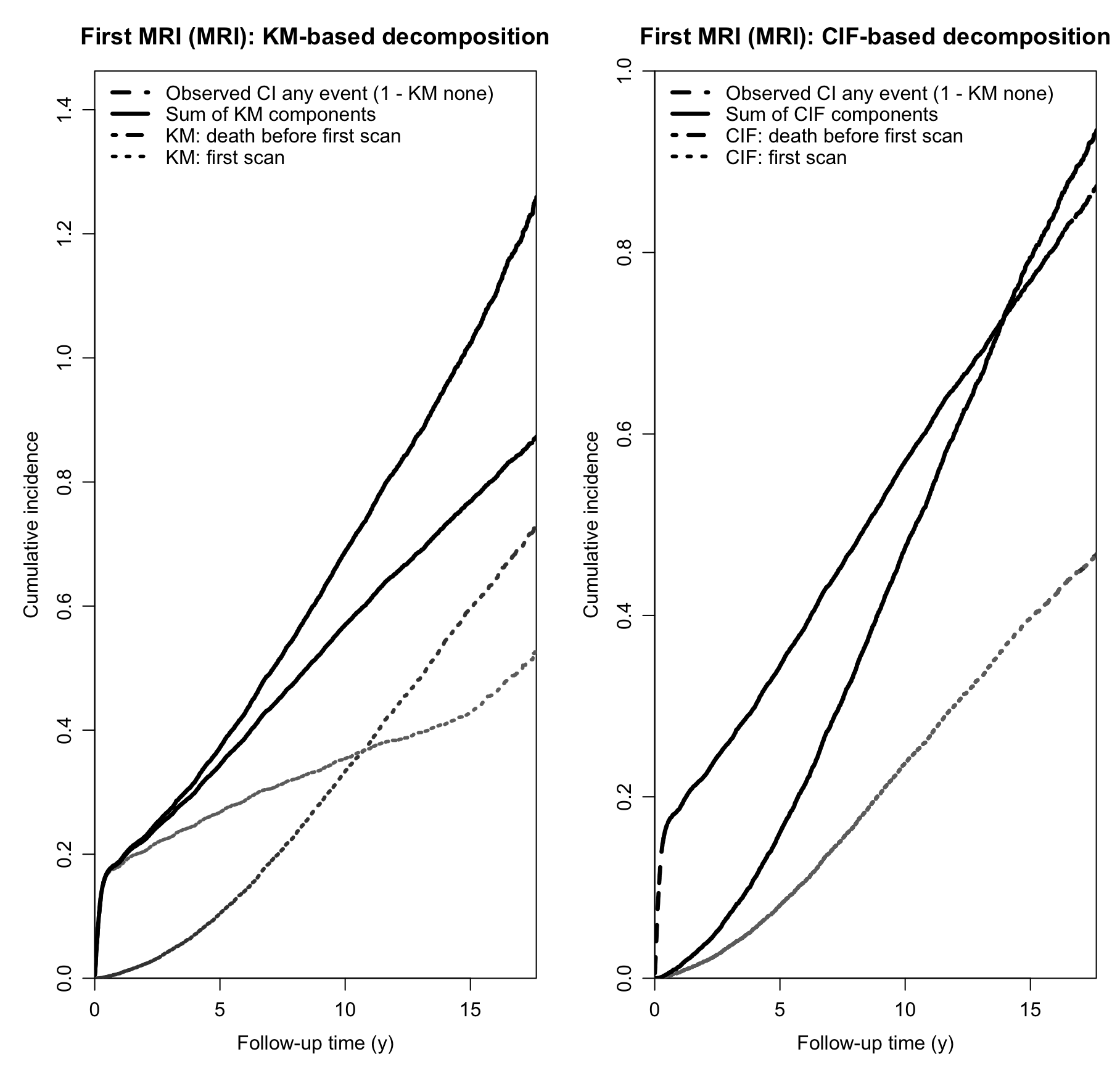}
\caption{Competing-risk summaries for first post-baseline MRI (Cause 1) and death before MRI (Cause 2).
Left panel: a na\"ive Kaplan-Meier-based decomposition that treats the competing event as independent censoring within each cause; because it combines two separate censoring assumptions, the resulting quantities are not coherent competing-risk probabilities and may exceed 1.
Right panel: cumulative incidence functions (CIFs), which provide coherent cause-specific event probabilities under competing risks and remain bounded in $[0,1]$.}
\label{fig:cif}
\end{figure}

The Figure~\ref{fig:cif} presents empirical summaries of the competing event processes for first post-baseline MRI (Cause~1) and death before MRI (Cause~2). The left panel shows a naïve Kaplan-Meier-based decomposition that treats the competing event as independent censoring within each cause. Because this approach combines incompatible censoring assumptions across causes, the resulting quantities do not represent coherent competing-risk probabilities and may exceed one.

The right panel displays cumulative incidence functions (CIFs), which provide valid cause-specific event probabilities under competing risks and remain bounded within the unit interval. Death accounted for a substantial fraction of total events over follow-up, particularly among older and more functionally impaired participants. These summaries demonstrate that treating death as benign censoring would overstate the probability of neuroimaging and motivate the use of competing-risk–aware regression models for inference \cite{KalbfleischPrentice2011,Beyersmann2012}.

\subsection{Primary analysis}
\subsubsection{Bayesian multilevel cause-specific AFT models}

Primary results from the Bayesian multilevel cause-specific accelerated failure time models are reported in Table~\ref{tab:bayes_main}. Models were adjusted for APOE $\varepsilon4$ status, age group, education, sex, race, baseline functional impairment, living situation, and vitamin B12 deficiency, and included an ADC-specific random intercept to quantify between-center differences in baseline timing. Results are presented as time ratios (TRs) with 95\% credible intervals, where values greater than one indicate longer time to the event and values less than one indicate earlier occurrence.
\begin{table}[!ht]
\centering
\caption{Bayesian accelerated failure time models for time to first MRI (Cause 1) and time to death before MRI (Cause 2).}
\label{tab:bayes_main}
\scriptsize
\begin{threeparttable}

\setlength{\tabcolsep}{6pt}
\renewcommand{\arraystretch}{1.15}

\begin{tabular}{>{\columncolor{tblbg}}l
                >{\columncolor{tblbg}}c
                >{\columncolor{tblbg}}c
                >{\columncolor{tblbg}}c
                >{\columncolor{tblbg}}c}
\toprule
\rowcolor{tblhead}
\textbf{Risk Factors} &
\multicolumn{2}{c}{\textbf{First MRI (Cause 1)}} &
\multicolumn{2}{c}{\textbf{Death before MRI (Cause 2)}} \\
\rowcolor{tblhead}
\cmidrule(lr){2-3} \cmidrule(lr){4-5}
 & \textbf{TR} & \textbf{95\% CrI} & \textbf{TR} & \textbf{95\% CrI} \\
\midrule

APOE $\varepsilon4$: One vs None
 & 0.98 & 0.85 to 1.13
 & 0.98 & 0.95 to 1.02 \\

APOE $\varepsilon4$: Two vs None
 & 0.79 & 0.60 to 1.06
 & 0.89\textsuperscript{*} & 0.82 to 0.97 \\[2pt]

Age 65--79 vs $<$65
 & 0.99 & 0.85 to 1.15
 & 0.69\textsuperscript{*} & 0.64 to 0.73 \\

Age $\ge$80 vs $<$65
 & 2.77\textsuperscript{*} & 2.20 to 3.51
 & 0.38\textsuperscript{*} & 0.36 to 0.41 \\[2pt]

Undergraduate vs $\le$HS
 & 0.74\textsuperscript{*} & 0.61 to 0.90
 & 1.09\textsuperscript{*} & 1.03 to 1.15 \\

Graduate vs $\le$HS
 & 0.63\textsuperscript{*} & 0.52 to 0.77
 & 1.18\textsuperscript{*} & 1.12 to 1.25 \\[2pt]

Black vs White
 & 1.66\textsuperscript{*} & 1.11 to 2.50
 & 0.86 & 0.71 to 1.04 \\

Other vs White
 & 1.86\textsuperscript{*} & 1.47 to 2.35
 & 0.69\textsuperscript{*} & 0.60 to 0.78 \\[2pt]

Female vs Male
 & 1.09 & 0.95 to 1.24
 & 1.25\textsuperscript{*} & 1.20 to 1.30 \\[2pt]

CDR Global $\ge$1 vs 0
 & 1.15 & 1.00 to 1.31
 & 1.52\textsuperscript{*} & 1.47 to 1.58 \\[2pt]

Live with spouse vs alone
 & 0.92 & 0.79 to 1.06
 & 1.09\textsuperscript{*} & 1.05 to 1.14 \\

Other household vs alone
 & 0.70\textsuperscript{*} & 0.56 to 0.87
 & 0.90\textsuperscript{*} & 0.84 to 0.97 \\[2pt]

Vitamin B12 deficiency (Yes vs No)
 & 0.91 & 0.71 to 1.18
 & 0.91\textsuperscript{*} & 0.84 to 0.99 \\

\bottomrule
\end{tabular}

\begin{tablenotes}[flushleft]\scriptsize
\item TR denotes time ratio from Bayesian accelerated failure time models. Values greater than 1 indicate longer time to event, and values less than 1 indicate shorter time to event, relative to the reference category. Cause 1 models time to first post-baseline MRI using a Weibull AFT specification, and Cause 2 models time to death before MRI using a log-logistic AFT specification. Credible intervals are 95 percent posterior intervals.
\item \textsuperscript{*} indicates the 95\% credible interval does not include 1. (HS: High School).
\end{tablenotes}

\end{threeparttable}
\end{table}

Across both event processes, APOE $\varepsilon4$ genotype showed divergent associations. APOE $\varepsilon4$ carriage was not credibly associated with the timing of first MRI. Relative to participants with no $\varepsilon4$ allele, one copy had a TR of 0.98 (95\% CrI 0.85 to 1.13), and two copies had a TR of 0.79 (0.60 to 1.06), indicating no clear acceleration or delay in receipt of neuroimaging after adjustment for demographic, clinical, and social factors. In contrast, APOE $\varepsilon4$ homozygosity was associated with earlier death before MRI (TR 0.89, 0.82 to 0.97), whereas one copy remained near null (TR 0.98, 0.95 to 1.02). This pattern indicates that genotype primarily influenced cohort composition through competing mortality rather than through differential imaging access.

Age was a dominant determinant of both processes. For MRI timing, participants aged 65 to 79 years were near null relative to those under 65 (TR 0.99, 0.85 to 1.15), whereas participants aged 80 years or older experienced substantially delayed imaging (TR 2.77, 2.20 to 3.51). For death before MRI, age effects were strong and monotone, with TRs of 0.69 (0.64 to 0.73) for ages 65 to 79 and 0.38 (0.36 to 0.41) for ages 80 and above, indicating markedly shorter time to death prior to imaging at older ages.

Educational attainment showed consistent and graded associations across both causes. Compared with participants with high school education or less, those with undergraduate education experienced earlier MRI (TR 0.74, 0.61 to 0.90) and longer time to death before MRI (TR 1.09, 1.03 to 1.15). These associations were stronger among participants with graduate education, who had TRs of 0.63 (0.52 to 0.77) for MRI timing and 1.18 (1.12 to 1.25) for death before MRI.

Racial differences were pronounced for MRI timing. Relative to White participants, Black participants experienced delayed imaging (TR 1.66, 1.11 to 2.50), as did participants in the Other race category (TR 1.86, 1.47 to 2.35). These disparities persisted after adjustment for education and baseline functional status. In contrast, race showed a different pattern for competing mortality: Black participants were near null (TR 0.86, 0.71 to 1.04), whereas participants in the Other race category experienced earlier death before MRI (TR 0.69, 0.60 to 0.78).

Sex and functional status showed outcome-specific associations. Female sex was near null for MRI timing (TR 1.09, 0.95 to 1.24) but was associated with longer time to death before MRI (TR 1.25, 1.20 to 1.30). Baseline functional impairment was weakly associated with delayed MRI (TR 1.15, 1.00 to 1.31) and strongly associated with death before MRI (TR 1.52, 1.47 to 1.58). This direction reflects the fitted AFT parameterization for Cause~2 and should be interpreted as a multiplicative effect on modeled time under the log-logistic specification.

Living arrangement and vitamin B12 deficiency showed more modest associations. Relative to living alone, living in another household arrangement was associated with earlier MRI (TR 0.70, 0.56 to 0.87), while living with a spouse or partner was near null. For death before MRI, living with a spouse or partner was associated with longer time (TR 1.09, 1.05 to 1.14), whereas other household arrangements were associated with shorter time (TR 0.90, 0.84 to 0.97). Vitamin B12 deficiency was not credibly associated with MRI timing (TR 0.91, 0.71 to 1.18) but was associated with earlier death before MRI (TR 0.91, 0.84 to 0.99).

\subsubsection{Center-level heterogeneity}

\begin{figure}[!ht]
\centering
\includegraphics[width=0.85\textwidth]{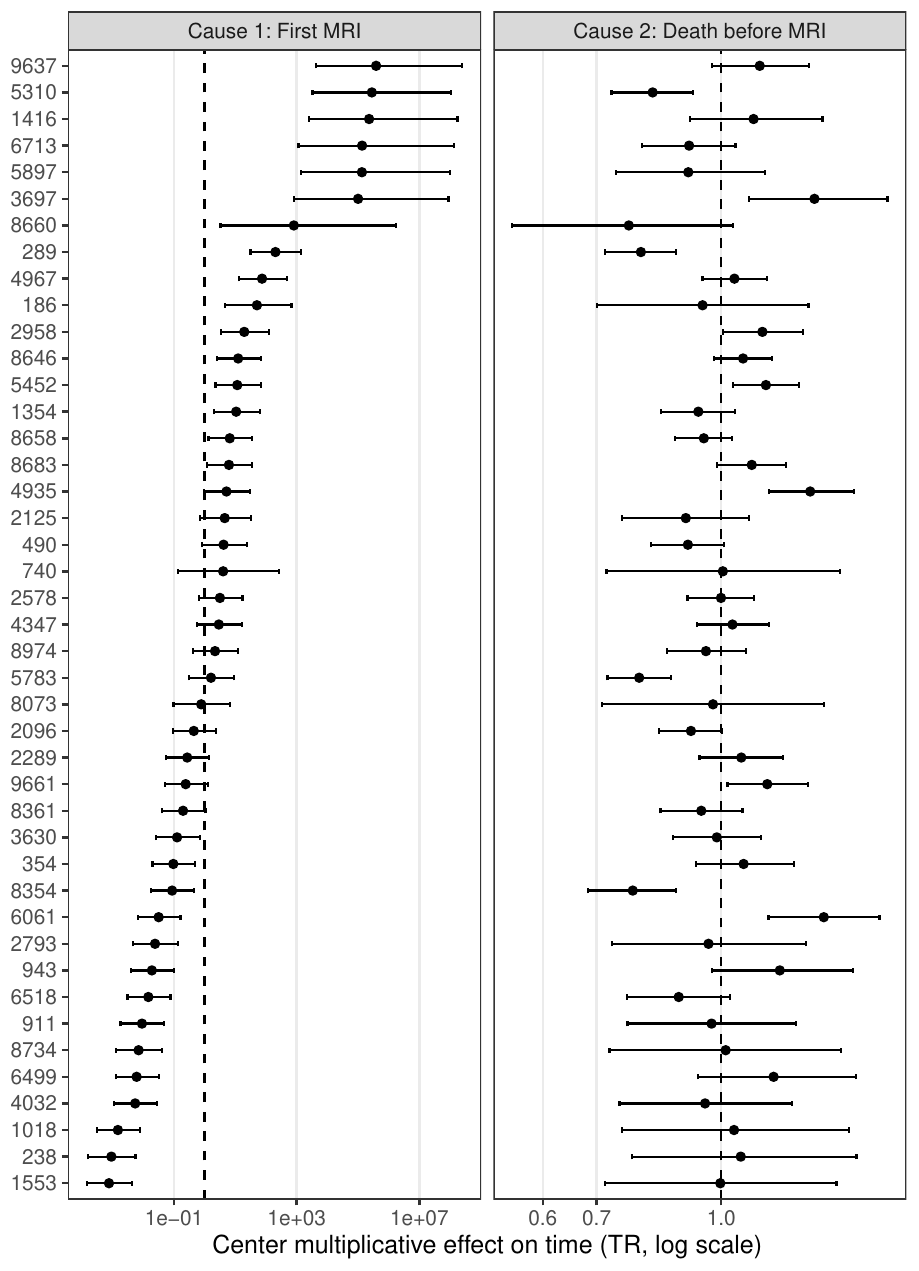}
\caption{Center-specific multiplicative effects on event timing from Bayesian AFT models. Each point represents the posterior mean time ratio (TR) for an Alzheimer’s Disease Center, with horizontal bars indicating 95\% credible intervals. The vertical dashed line denotes the overall average center effect (TR = 1). Left: time to first MRI. Right: time to death before MRI.}
\label{fig:caterpillar}
\end{figure}

Between-center heterogeneity in event timing was evaluated using the posterior distributions of the center-specific random intercepts from the Bayesian multilevel AFT models. Figure~\ref{fig:caterpillar} displays caterpillar plots of the estimated center-level multiplicative effects on time, expressed as time ratios (TR) on a logarithmic scale, for (A) time to first post-baseline MRI and (B) time to death before MRI. Each point represents the posterior mean for an individual Alzheimer’s Disease Center, with horizontal bars denoting 95\% credible intervals. The vertical reference line at TR = 1 corresponds to the average center after adjustment for individual-level covariates.
Marked heterogeneity across centers was observed for the timing of first MRI. Center-specific effects spanned a wide range, with several centers exhibiting substantially delayed imaging relative to the overall average, while others demonstrated meaningfully accelerated access. This variability persisted despite adjustment for age, education, race, cognitive status, living situation, and APOE $\varepsilon4$ genotype. Consistent with these visual patterns, the posterior mean standard deviation of the center-level random intercepts for Cause~1 was 5.62 (95\% CrI 4.41 to 7.24), indicating large structural differences across centers in baseline imaging timing on the log-time scale.

Whereas, center-level variability for death before MRI was considerably more limited. The caterpillar plot for Cause~2 shows center-specific effects tightly clustered around the null, with substantially narrower credible intervals. The posterior mean standard deviation of the center-level random intercepts for Cause~2 was 0.17 (95\% CrI 0.12 to 0.23), indicating modest between-center differences in mortality timing after adjustment for individual-level characteristics.

These results demonstrate that institutional factors play a dominant role in shaping access to neuroimaging but contribute far less to short-term mortality risk within this cohort. The divergence in heterogeneity across the two causes indicates that center-level effects primarily operate through diagnostic and operational pathways rather than through differences in underlying survival. These findings directly support the study objective of quantifying institutional contributions to disparities in neuroimaging timing and underscore the importance of explicitly modeling site effects when imaging availability defines cohort inclusion.

\subsection{Bayesian MCMC diagnostics}

Markov chain Monte Carlo diagnostics indicated satisfactory convergence and mixing for both cause-specific Bayesian models. Trace plots, autocorrelation functions, and posterior density overlays showed stable sampling behavior and are presented in Figures~\ref{fig:aft-mri_acf}, \ref{fig:aft-death_acf}, \ref{fig:aft-mri_trace},\ref{fig:aft-death_trace}, \ref{fig:aft-mri_over}, and \ref{fig:aft-death_over}. Split-chain $\widehat{R}$ statistics were close to 1.00 for all regression coefficients and variance parameters, and effective sample sizes were large relative to the total number of iterations. These diagnostics support the reliability of posterior summaries reported in the primary analyses.

\subsection{Sensitivity analyses}

Sensitivity analyses were conducted to assess robustness to modeling assumptions and inferential framework. Frequentist cause-specific AFT models yielded effect directions and relative magnitudes consistent with the Bayesian results for both causes, with age, education, and race emerging as the dominant determinants of MRI timing and competing mortality (Table~\ref{tab:freq_aft_combined}).
\begin{table*}[!ht]
\centering
\caption{Frequentist accelerated failure time models as sensitivity analyses for time to first MRI (Cause 1) and time to death before MRI (Cause 2).}
\label{tab:freq_aft_combined}
\scriptsize
\begin{threeparttable}
\setlength{\tabcolsep}{6pt}
\renewcommand{\arraystretch}{1.15}

\begin{tabular}{>{\columncolor{tblbg}}l
                >{\columncolor{tblbg}}c
                >{\columncolor{tblbg}}c
                >{\columncolor{tblbg}}c
                >{\columncolor{tblbg}}c}
\toprule
\rowcolor{tblhead}
\textbf{Risk Factors} &
\multicolumn{2}{c}{\textbf{First MRI (Cause 1)}} &
\multicolumn{2}{c}{\textbf{Death before MRI (Cause 2)}} \\
\rowcolor{tblhead}
\cmidrule(lr){2-3} \cmidrule(lr){4-5}
 & \textbf{TR} & \textbf{95\% CI} & \textbf{TR} & \textbf{95\% CI} \\
\midrule
APOE $\varepsilon4$: One vs None
 & 0.964 & 0.814 to 1.141
 & 0.977 & 0.940 to 1.016 \\

APOE $\varepsilon4$: Two vs None
 & 0.927 & 0.646 to 1.328
 & 0.868\textsuperscript{*} & 0.797 to 0.945 \\[4pt]

Age 65--79 vs $<$65
 & 1.976\textsuperscript{*} & 1.648 to 2.370
 & 0.690\textsuperscript{*} & 0.648 to 0.735 \\

Age $\ge$80 vs $<$65
 & 9.859\textsuperscript{*} & 7.418 to 13.104
 & 0.376\textsuperscript{*} & 0.351 to 0.402 \\[4pt]

College or undergraduate ($\le$16 y) vs $\le$HS
 & 0.588\textsuperscript{*} & 0.461 to 0.748
 & 1.082\textsuperscript{*} & 1.028 to 1.138 \\

Graduate ($>$16 y) vs $\le$HS
 & 0.416\textsuperscript{*} & 0.326 to 0.531
 & 1.190\textsuperscript{*} & 1.129 to 1.254 \\[4pt]

Black vs White
 & 4.558\textsuperscript{*} & 2.708 to 7.671
 & 0.799\textsuperscript{*} & 0.663 to 0.963 \\

Other vs White
 & 4.531\textsuperscript{*} & 3.409 to 6.022
 & 0.663\textsuperscript{*} & 0.583 to 0.754 \\[4pt]

Female vs Male
 & 0.929 & 0.787 to 1.098
 & 1.250\textsuperscript{*} & 1.202 to 1.300 \\[4pt]

CDR Global $\ge$1 vs 0
 & 1.209\textsuperscript{*} & 1.025 to 1.427
 & 1.505\textsuperscript{*} & 1.450 to 1.562 \\[4pt]

Live with spouse vs alone
 & 0.995 & 0.825 to 1.200
 & 1.082\textsuperscript{*} & 1.037 to 1.130 \\

Other household vs alone
 & 0.486\textsuperscript{*} & 0.366 to 0.646
 & 0.894\textsuperscript{*} & 0.830 to 0.962 \\[4pt]

Vitamin B12 deficiency (Yes vs No)
 & 0.557\textsuperscript{*} & 0.404 to 0.769
 & 0.890\textsuperscript{*} & 0.822 to 0.964 \\

\bottomrule
\end{tabular}

\begin{tablenotes}[flushleft]\scriptsize
\item TR denotes time ratio from frequentist accelerated failure time models. TR greater than 1 indicates longer time to the event (delay), and TR less than 1 indicates shorter time to the event (acceleration), relative to the reference category. Cause 1 uses a Weibull AFT model for time to first MRI. Cause 2 uses a log-logistic AFT model for time to death before MRI. All models adjust for the covariates shown. \textsuperscript{*} Indicates 95\% confidence interval does not include 1.
\end{tablenotes}
\end{threeparttable}
\end{table*}

Cause-specific Cox models for time to first MRI showed evidence of nonproportional hazards for several key covariates based on Schoenfeld residual diagnostics. This motivated the use of time-varying coefficient models. Extended Cox models incorporating interactions with log follow-up time indicated that disparities in imaging access by age, education, and race were most pronounced early after baseline and attenuated over follow-up. Figure~\ref{fig:tvc} displays the estimated time-varying hazard ratios for age groups, education, and race relative to their reference categories.

For age, older participants exhibited a higher hazard of receiving MRI early in follow-up, with hazard ratios declining over time and approaching or crossing unity by approximately 10-15 years. Higher educational attainment was associated with a substantially elevated early hazard of MRI, which diminished with longer follow-up. Black and Other race participants consistently showed a lower hazard of MRI than White participants, with the largest disparities observed early after baseline. In contrast, APOE $\varepsilon4$ genotype was not associated with the timing of first MRI in any sensitivity analysis.

These hazard-based findings align with the Bayesian AFT results and support the interpretation that observed disparities reflect differences in early access to neuroimaging rather than persistent differences in long-term imaging rates. Full sensitivity analyses, including distributional diagnostics, model comparison metrics, and time-varying hazard ratio curves, are provided in the Supplementary Materials (\textbf{Table S4}).

\section{Discussion}

Brain imaging in Alzheimer's disease research occurs within a real clinical world, not as a passive prerequisite for downstream analysis. When and whether someone receives an MRI reflects their biological risk, social circumstances, and the organization of the healthcare system around them. In this study, we examined when participants in the National Alzheimer's Coordinating Center (NACC) dataset received their first brain MRI after enrollment. We treated death before imaging as a competing outcome that prevents scanning from ever occurring, rather than as non-informative censoring. This matters because in older adults, death is common and unevenly distributed. It connects to the same factors that influence who gets imaging and when. Our approach acknowledges that diagnostic procedures unfold in real clinical settings, shaped by more than research protocols alone.

Two contributions motivate the analysis. First, we quantify individual-level determinants of MRI timing and competing mortality using time ratios that directly measure delay or acceleration on the time scale. Second, we characterize center-level heterogeneity in neuroimaging timing across Alzheimer's Disease Centers (ADCs), an aspect with direct implications for equity, cohort composition, and the interpretability of imaging-based AD research.

A central finding is that APOE $\varepsilon4$ genotype was not meaningfully associated with the timing of first MRI after adjustment for demographic, cognitive, and social factors. For example, APOE $\varepsilon4$ homozygosity was associated with a time ratio of approximately 0.8 for MRI timing, with 95\% credible intervals spanning the null, indicating no clear acceleration or delay in imaging. This result stands in contrast to the well-established role of APOE $\varepsilon4$ in AD risk, biomarker progression, and clinical decline \cite{Liu2013,Jack2018}. The absence of a strong genotype effect on imaging timing suggests that, within ADC settings, neuroimaging decisions are driven less by genetic risk stratification and more by clinical presentation, functional impairment, and structural features of care delivery.

Conversely, APOE $\varepsilon4$ homozygosity was associated with earlier death prior to imaging, with time ratios consistently below 1 and credible intervals excluding the null. This distinction highlights the importance of separating the imaging and mortality processes. Genetic risk appears to exert greater influence on survival than on access to diagnostic imaging, yet genotype-linked mortality can indirectly shape imaging cohorts by removing high-risk individuals before imaging occurs. Without explicit competing risk modeling, this selection mechanism would remain hidden, potentially biasing inferences drawn from imaging-based samples.

Age, education, and race emerged as the dominant determinants of MRI timing. Participants aged 80 years or older experienced substantial delays in imaging, with time ratios exceeding 2 relative to participants younger than 65, even after adjustment for cognitive status and functional impairment. Educational attainment showed a graded association with earlier imaging; participants with post-secondary education underwent MRI substantially sooner than those with a high school education or less. These patterns are consistent with prior evidence linking education to health system navigation, continuity of care, and engagement with diagnostic services.

Racial disparities persisted after multivariable adjustment. Black participants and those categorized as other race experienced delayed imaging relative to White participants, with time ratios well above 1 for MRI timing. These findings align with extensive literature documenting racial disparities in neurologic care utilization and access to advanced neuroimaging in the United States \cite{Williams1997,Braveman2011,Saadi2017,Vagal2023,Lim2023}. Importantly, these disparities were observed within a cohort enriched for specialty care and research participation, indicating that inclusion in ADC networks does not fully eliminate inequities in diagnostic pathways. This has direct implications for the representativeness of imaging-based AD research and for efforts to ensure equitable access to emerging diagnostic and therapeutic tools.

The competing mortality process clarifies why competing risk methods are essential when neuroimaging timing is the outcome of interest. Older age and greater functional impairment were associated both with delayed imaging and with increased risk of death prior to imaging. Treating death as simple censoring in this setting would overstate the cumulative incidence of MRI and misattribute differences in imaging timing to access alone, when mortality plays a concurrent and differential role \cite{Pintilie2006,Beyersmann2012,KalbfleischPrentice2011}. By modeling imaging and death separately, we demonstrate that their determinants overlap but are not identical, and that valid inference on imaging access requires explicit consideration of competing mortality.

A distinctive contribution of this work is the magnitude of center-level heterogeneity in MRI timing.  After adjustment for individual-level covariates, substantial between-center variability remained for time to first MRI, with random-intercept standard deviations indicating large differences in baseline imaging timing across ADCs. By comparison, center-level variability was much smaller for death prior to MRI. This asymmetry suggests that diagnostic procedures are particularly sensitive to center-specific capacity, referral patterns, scheduling practices, and operational workflows, whereas mortality processes are less dependent on site-level factors. These findings align with recent documentation of substantial site-level variability in participant recruitment profiles and baseline characteristics across NACC centers \cite{Chan2025}. From a research design perspective, this finding is critical. Studies that condition on receipt of MRI may inherit center-specific selection pressures, potentially reducing transportability and generating apparent differences in imaging biomarkers that reflect access rather than underlying disease biology.

Methodologically, the accelerated failure time framework aligns closely with the scientific goal of quantifying delay and acceleration in diagnostic processes. Time ratios provide direct, clinically interpretable measures of access and timing, and avoid reliance on proportional hazards assumptions that were unstable for several key covariates in this cohort. Sensitivity analyses using time-varying Cox models supported this assessment and suggested that disparities in imaging access may be most pronounced early after cohort entry. Bayesian hierarchical modeling provided coherent uncertainty quantification and a principled approach to estimating center-level heterogeneity while retaining interpretable covariate effects \cite{Gelman2013,Buerkner2017,Carpenter2017}. Importantly, prior sensitivity analyses demonstrated that substantive conclusions were robust across a wide range of plausible prior specifications, indicating that results were driven by the observed data rather than by modeling choices.

Several limitations merit consideration. NACC does not capture detailed information on imaging indications, insurance coverage, geographic proximity to imaging facilities, or center-specific imaging protocols, leaving open the possibility of residual confounding by unmeasured structural factors. The random-intercept specification captures baseline center-level differences but does not identify specific operational mechanisms or allow center-specific covariate effects. Additionally, covariates were defined at baseline and do not capture time-updated changes in cognition, living situation, or clinical status that may influence imaging decisions. Finally, because the cohort is drawn from ADC settings, findings may not generalize directly to community-based care environments, where access barriers may be greater.

Even with these limitations, the results have clear implications for equity and inference in Alzheimer's disease research. Imaging-based studies implicitly condition on neuroimaging availability, yet our findings show that both the timing and occurrence of MRI are socially patterned and center-dependent. Without explicit competing risk and site-aware modeling, imaging-based cohorts may underrepresent older, less educated, and non-White individuals, not due to lower disease burden, but due to delayed or foregone imaging combined with competing mortality. Addressing this challenge requires both analytic strategies that respect the competing risk structure and structural efforts to standardize and improve equitable neuroimaging workflows across centers.

\section{Conclusion}
In this large multicenter cohort of NACC participants without dementia at baseline, the timing of first brain MRI after enrollment was shaped primarily by demographic, social, and clinical factors rather than by APOE $\varepsilon4$ genotype. APOE $\varepsilon4$ homozygosity was associated with earlier death before imaging could occur, but not with faster or slower access to MRI itself. Competing mortality played a substantial role in determining who received imaging and when, confirming that death cannot be treated as non-informative censoring when neuroimaging timing is the outcome. Bayesian multilevel modeling revealed considerable variability across Alzheimer's Disease Centers in MRI timing, indicating that access to neuroimaging depends heavily on site-specific factors even within a coordinated national research network. These findings have direct implications for how imaging-based studies are designed and interpreted. They point to the need for equity-focused neuroimaging workflows, careful attention to center effects in cohort construction, and routine use of competing risk methods when analyzing time-to-imaging data in older populations.

\section*{Data Availability Statement}
The NACC database is funded by NIA/NIH Grant U24 AG072122. SCAN is a multi-institutional project that was funded as a U24 grant (AG067418) by the National Institute on Aging in May 2020. Data collected by SCAN and shared by NACC are contributed by the NIA-funded ADRCs as follows:

Arizona Alzheimer’s Center - P30 AG072980 (PI: Eric Reiman, MD); R01 AG069453 (PI: Eric Reiman (contact), MD); P30 AG019610 (PI: Eric Reiman, MD); and the State of Arizona which provided additional funding supporting our center; Boston University - P30 AG013846 (PI Neil Kowall MD); Cleveland ADRC - P30 AG062428 (James Leverenz, MD); Cleveland Clinic, Las Vegas – P20AG068053; Columbia - P50 AG008702 (PI Scott Small MD); Duke/UNC ADRC – P30 AG072958; Emory University - P30AG066511 (PI Levey Allan, MD, PhD); Indiana University - R01 AG19771 (PI Andrew Saykin, PsyD); P30 AG10133 (PI Andrew Saykin, PsyD); P30 AG072976 (PI Andrew Saykin, PsyD); R01 AG061788 (PI Shannon Risacher, PhD); R01 AG053993 (PI Yu-Chien Wu, MD, PhD); U01 AG057195 (PI Liana Apostolova, MD); U19 AG063911 (PI Bradley Boeve, MD); and the Indiana University Department of Radiology and Imaging Sciences; Johns Hopkins - P30 AG066507 (PI Marilyn Albert, Phd.); Mayo Clinic - P50 AG016574 (PI Ronald Petersen MD PhD); Mount Sinai - P30 AG066514 (PI Mary Sano, PhD); R01 AG054110 (PI Trey Hedden, PhD); R01 AG053509 (PI Trey Hedden, PhD); New York University - P30AG066512-01S2 (PI Thomas Wisniewski, MD); R01AG056031 (PI Ricardo Osorio, MD); R01AG056531 (PIs Ricardo Osorio, MD; Girardin Jean-Louis, PhD); Northwestern University - P30 AG013854 (PI Robert Vassar PhD); R01 AG045571 (PI Emily Rogalski, PhD); R56 AG045571, (PI Emily Rogalski, PhD); R01 AG067781, (PI Emily Rogalski, PhD); U19 AG073153, (PI Emily Rogalski, PhD); R01 DC008552, (M.-Marsel Mesulam, MD); R01 AG077444, (PIs M.-Marsel Mesulam, MD, Emily Rogalski, PhD); R01 NS075075 (PI Emily Rogalski, PhD); R01 AG056258 (PI Emily Rogalski, PhD); Oregon Health and Science University - P30 AG066518 (PIs Lisa Silbert, MD, Kevin Duff, PhD); R56 AG074321 (PI Jeffrey Kaye, MD); Rush University - P30 AG010161 (PI David Bennett MD); Stanford – P30AG066515; P50 AG047366 (PI Victor Henderson MD MS); University of Alabama, Birmingham – P20; University of California, Davis - P30 AG10129 (PI Charles DeCarli, MD); P30 AG072972 (PI Charles DeCarli, MD); University of California, Irvine - P50 AG016573 (PI Frank LaFerla PhD); University of California, San Diego - P30AG062429 (PI James Brewer, MD, PhD); University of California, San Francisco - P30 AG062422 (Rabinovici, Gil D., MD); University of Kansas - P30 AG035982 (Russell Swerdlow, MD); University of Kentucky - P30 AG028283-15S1 (PIs Linda Van Eldik, PhD and Brian Gold, PhD); University of Michigan ADRC - P30AG053760 (PI Henry Paulson, MD, PhD) P30AG072931 (PI Henry Paulson, MD, PhD) Cure Alzheimer's Fund 200775 - (PI Henry Paulson, MD, PhD) U19 NS120384 (PI Charles DeCarli, MD, University of Michigan Site PI Henry Paulson, MD, PhD) R01 AG068338 (MPI Bruno Giordani, PhD, Carol Persad, PhD, Yi Murphey, PhD) S10OD026738-01 (PI Douglas Noll, PhD) R01 AG058724 (PI Benjamin Hampstead, PhD) R35 AG072262 (PI Benjamin Hampstead, PhD) W81XWH2110743 (PI Benjamin Hampstead, PhD) R01 AG073235 (PI Nancy Chiaravalloti, University of Michigan Site PI Benjamin Hampstead, PhD) 1I01RX001534 (PI Benjamin Hampstead, PhD) IRX001381 (PI Benjamin Hampstead, PhD); University of New Mexico - P20 AG068077 (Gary Rosenberg, MD); University of Pennsylvania - State of PA project 2019NF4100087335 (PI David Wolk, MD); Rooney Family Research Fund (PI David Wolk, MD); R01 AG055005 (PI David Wolk, MD); University of Pittsburgh - P50 AG005133 (PI Oscar Lopez MD); University of Southern California - P50 AG005142 (PI Helena Chui MD); University of Washington - P50 AG005136 (PI Thomas Grabowski MD); University of Wisconsin - P50 AG033514 (PI Sanjay Asthana MD FRCP); Vanderbilt University – P20 AG068082; Wake Forest - P30AG072947 (PI Suzanne Craft, PhD); Washington University, St. Louis - P01 AG03991 (PI John Morris MD); P01 AG026276 (PI John Morris MD); P20 MH071616 (PI Dan Marcus); P30 AG066444 (PI John Morris MD); P30 NS098577 (PI Dan Marcus); R01 AG021910 (PI Randy Buckner); R01 AG043434 (PI Catherine Roe); R01 EB009352 (PI Dan Marcus); UL1 TR000448 (PI Brad Evanoff); U24 RR021382 (PI Bruce Rosen); Avid Radiopharmaceuticals / Eli Lilly; Yale - P50 AG047270 (PI Stephen Strittmatter MD PhD); R01AG052560 (MPI: Christopher van Dyck, MD; Richard Carson, PhD); R01AG062276 (PI: Christopher van Dyck, MD); 1Florida - P30AG066506-03 (PI Glenn Smith, PhD); P50 AG047266 (PI Todd Golde MD PhD) 

\section{Funding}
This research received no specific grant from any funding agency in the public, commercial, or not-for-profit sectors.

\section{Competing Interest}
None declared.

\clearpage

\setcounter{section}{0}
\setcounter{figure}{0}
\setcounter{table}{0}

\renewcommand{\thesection}{S\arabic{section}}
\renewcommand{\thesubsection}{S\arabic{section}.\arabic{subsection}}
\renewcommand{\thefigure}{S\arabic{figure}}
\renewcommand{\thetable}{S\arabic{table}}

\section{Supplementary Material}
\subsection{Prior sensitivity analysis}

Because Bayesian hierarchical survival models can be sensitive to assumptions about the amount of shrinkage and tail behavior of priors on regression and variance components, we evaluated robustness of the primary conclusions under several alternative prior sets. These alternatives were designed to span: (i) heavier-tailed priors on positive parameters, (ii) stronger shrinkage on regression coefficients, (iii) alternative priors for center-level random-effect standard deviations, and (iv) broader priors on intercepts.

Table~\ref{tab:prior-sens-tr} summarizes posterior time ratios (TR) and 95\% credible intervals for both causes under each prior set. Across all prior choices, the direction and practical magnitude of key findings were stable: APOE $\varepsilon4$ showed no meaningful association with time to first MRI, while age, education, race, and center-level heterogeneity remained the dominant drivers of MRI timing. For death before MRI, the association of APOE $\varepsilon4$ homozygosity with shorter time to death and the strong age and impairment effects were also consistent across prior sets. Overall, the prior sensitivity results support that the main inferences are not artifacts of a single prior specification.

\clearpage
\begin{landscape}

\begin{table*}[!ht]
\centering
\caption{Prior sensitivity analysis for Bayesian multilevel cause-specific accelerated failure time (AFT) models: time ratios (TR) and 95\% credible intervals for time to first MRI (Cause~1) and death before MRI (Cause~2) under alternative prior specifications.}
\label{tab:prior-sens-tr}

\footnotesize
\setlength{\tabcolsep}{3.2pt}
\renewcommand{\arraystretch}{1.15}

\begin{threeparttable}

\begin{adjustbox}{max width=\linewidth, max totalheight=0.95\textheight, keepaspectratio}

\begin{tabular}{
>{\columncolor{tblbg}}l
>{\columncolor{tblbg}}c >{\columncolor{tblbg}}c
>{\columncolor{tblbg}}c >{\columncolor{tblbg}}c
>{\columncolor{tblbg}}c >{\columncolor{tblbg}}c
>{\columncolor{tblbg}}c >{\columncolor{tblbg}}c
>{\columncolor{tblbg}}c >{\columncolor{tblbg}}c
>{\columncolor{tblbg}}c >{\columncolor{tblbg}}c
>{\columncolor{tblbg}}c >{\columncolor{tblbg}}c
}
\toprule
\rowcolor{tblhead}
\textbf{Variable} &
\multicolumn{2}{c}{\textbf{Baseline (A)}} &
\multicolumn{2}{c}{\textbf{Heavy-tail (B)}} &
\multicolumn{2}{c}{\textbf{More shrinkage (C)}} &
\multicolumn{2}{c}{\textbf{Half-normal RE (D)}} &
\multicolumn{2}{c}{\textbf{Wide intercept (E)}} &
\multicolumn{2}{c}{\textbf{Strong slopes (F)}} &
\multicolumn{2}{c}{\textbf{Student-$t$ slopes (G)}} \\

\rowcolor{tblhead}
 &
\textbf{C1} & \textbf{C2} &
\textbf{C1} & \textbf{C2} &
\textbf{C1} & \textbf{C2} &
\textbf{C1} & \textbf{C2} &
\textbf{C1} & \textbf{C2} &
\textbf{C1} & \textbf{C2} &
\textbf{C1} & \textbf{C2} \\

\cmidrule(lr){2-3}\cmidrule(lr){4-5}\cmidrule(lr){6-7}
\cmidrule(lr){8-9}\cmidrule(lr){10-11}\cmidrule(lr){12-13}\cmidrule(lr){14-15}
\midrule

APOE $\varepsilon4$: one copy vs none
& 0.98 (0.85--1.13) & 0.99 (0.96--1.03)
& 0.98 (0.85--1.14) & 0.99 (0.96--1.03)
& 0.98 (0.85--1.13) & 0.99 (0.96--1.03)
& 0.98 (0.85--1.13) & 0.99 (0.96--1.03)
& 0.98 (0.85--1.13) & 0.99 (0.96--1.03)
& 0.98 (0.85--1.13) & 0.99 (0.96--1.03)
& 0.98 (0.85--1.13) & 0.99 (0.96--1.03) \\

APOE $\varepsilon4$: two copies vs none
& 0.79 (0.60--1.06) & 0.89 (0.83--0.97)
& 0.79 (0.60--1.06) & 0.89 (0.83--0.97)
& 0.79 (0.60--1.06) & 0.89 (0.83--0.97)
& 0.79 (0.60--1.06) & 0.89 (0.83--0.97)
& 0.79 (0.60--1.06) & 0.89 (0.83--0.97)
& 0.79 (0.60--1.06) & 0.89 (0.83--0.97)
& 0.79 (0.60--1.06) & 0.89 (0.83--0.97) \\

Age 65--79 vs $<65$
& 1.00 (0.86--1.17) & 0.69 (0.65--0.74)
& 1.00 (0.86--1.17) & 0.69 (0.65--0.74)
& 1.00 (0.86--1.17) & 0.69 (0.65--0.74)
& 1.00 (0.86--1.17) & 0.69 (0.65--0.74)
& 1.00 (0.86--1.17) & 0.69 (0.65--0.74)
& 1.00 (0.86--1.17) & 0.69 (0.65--0.74)
& 1.00 (0.86--1.17) & 0.69 (0.65--0.74) \\

Age $\ge 80$ vs $<65$
& 2.79 (2.20--3.54) & 0.38 (0.36--0.41)
& 2.79 (2.20--3.55) & 0.38 (0.36--0.41)
& 2.79 (2.20--3.54) & 0.38 (0.36--0.41)
& 2.79 (2.20--3.54) & 0.38 (0.36--0.41)
& 2.79 (2.20--3.54) & 0.38 (0.36--0.41)
& 2.79 (2.20--3.54) & 0.38 (0.36--0.41)
& 2.79 (2.20--3.54) & 0.38 (0.36--0.41) \\

Education $\le16$y vs $\le$HS
& 0.74 (0.61--0.90) & 1.09 (1.03--1.15)
& 0.74 (0.61--0.90) & 1.09 (1.03--1.15)
& 0.74 (0.61--0.90) & 1.09 (1.03--1.15)
& 0.74 (0.61--0.90) & 1.09 (1.03--1.15)
& 0.74 (0.61--0.90) & 1.09 (1.03--1.15)
& 0.74 (0.61--0.90) & 1.09 (1.03--1.15)
& 0.74 (0.61--0.90) & 1.09 (1.03--1.15) \\

Education $>16$y vs $\le$HS
& 0.63 (0.52--0.77) & 1.18 (1.12--1.25)
& 0.63 (0.52--0.77) & 1.18 (1.12--1.25)
& 0.63 (0.52--0.77) & 1.18 (1.12--1.25)
& 0.63 (0.52--0.77) & 1.18 (1.12--1.25)
& 0.63 (0.52--0.77) & 1.18 (1.12--1.25)
& 0.63 (0.52--0.77) & 1.18 (1.12--1.25)
& 0.63 (0.52--0.77) & 1.18 (1.12--1.25) \\

Race: Black vs White
& 1.66 (1.11--2.50) & 0.86 (0.71--1.04)
& 1.66 (1.11--2.50) & 0.86 (0.71--1.04)
& 1.66 (1.11--2.50) & 0.86 (0.71--1.04)
& 1.66 (1.11--2.50) & 0.86 (0.71--1.04)
& 1.66 (1.11--2.50) & 0.86 (0.71--1.04)
& 1.66 (1.11--2.50) & 0.86 (0.71--1.04)
& 1.66 (1.11--2.50) & 0.86 (0.71--1.04) \\

Race: Other vs White
& 1.86 (1.47--2.35) & 0.69 (0.61--0.78)
& 1.86 (1.47--2.35) & 0.69 (0.61--0.78)
& 1.86 (1.47--2.35) & 0.69 (0.61--0.78)
& 1.86 (1.47--2.35) & 0.69 (0.61--0.78)
& 1.86 (1.47--2.35) & 0.69 (0.61--0.78)
& 1.86 (1.47--2.35) & 0.69 (0.61--0.78)
& 1.86 (1.47--2.35) & 0.69 (0.61--0.78) \\

Sex: Female vs Male
& 0.93 (0.82--1.06) & 1.25 (1.20--1.30)
& 0.93 (0.82--1.06) & 1.25 (1.20--1.30)
& 0.93 (0.82--1.06) & 1.25 (1.20--1.30)
& 0.93 (0.82--1.06) & 1.25 (1.20--1.30)
& 0.93 (0.82--1.06) & 1.25 (1.20--1.30)
& 0.93 (0.82--1.06) & 1.25 (1.20--1.30)
& 0.93 (0.82--1.06) & 1.25 (1.20--1.30) \\

CDR Global $\ge1$ vs 0
& 1.15 (1.00--1.32) & 1.52 (1.47--1.58)
& 1.15 (1.00--1.32) & 1.52 (1.47--1.58)
& 1.15 (1.00--1.32) & 1.52 (1.47--1.58)
& 1.15 (1.00--1.32) & 1.52 (1.47--1.58)
& 1.15 (1.00--1.32) & 1.52 (1.47--1.58)
& 1.15 (1.00--1.32) & 1.52 (1.47--1.58)
& 1.15 (1.00--1.32) & 1.52 (1.47--1.58) \\

Living with spouse vs alone
& 0.92 (0.79--1.06) & 1.09 (1.05--1.14)
& 0.92 (0.79--1.06) & 1.09 (1.05--1.14)
& 0.92 (0.79--1.06) & 1.09 (1.05--1.14)
& 0.92 (0.79--1.06) & 1.09 (1.05--1.14)
& 0.92 (0.79--1.06) & 1.09 (1.05--1.14)
& 0.92 (0.79--1.06) & 1.09 (1.05--1.14)
& 0.92 (0.79--1.06) & 1.09 (1.05--1.14) \\

Other household vs alone
& 0.70 (0.56--0.87) & 0.90 (0.84--0.97)
& 0.70 (0.56--0.87) & 0.90 (0.84--0.97)
& 0.70 (0.56--0.87) & 0.90 (0.84--0.97)
& 0.70 (0.56--0.87) & 0.90 (0.84--0.97)
& 0.70 (0.56--0.87) & 0.90 (0.84--0.97)
& 0.70 (0.56--0.87) & 0.90 (0.84--0.97)
& 0.70 (0.56--0.87) & 0.90 (0.84--0.97) \\

B12 deficiency vs none
& 0.91 (0.71--1.19) & 0.91 (0.84--0.99)
& 0.91 (0.71--1.19) & 0.91 (0.84--0.99)
& 0.91 (0.71--1.19) & 0.91 (0.84--0.99)
& 0.91 (0.71--1.19) & 0.91 (0.84--0.99)
& 0.91 (0.71--1.19) & 0.91 (0.84--0.99)
& 0.91 (0.71--1.19) & 0.91 (0.84--0.99)
& 0.91 (0.71--1.19) & 0.91 (0.84--0.99) \\

\bottomrule
\end{tabular}
\end{adjustbox}

\begin{tablenotes}[flushleft]\footnotesize
\item Entries are time ratios (TR) with 95\% credible intervals. TR $>$ 1 indicates longer time to event (delay), and TR $<$ 1 indicates shorter time to event (acceleration), relative to the \\reference category.
\item Cause~1 models time to first post-baseline MRI using a Weibull AFT specification. Cause~2 models time to death before MRI using a log-logistic AFT specification.
\item Reference categories: Age $<65$ years; Education $\le$ high school; Race White; Sex Male; Living situation alone; CDR Global = 0; APOE $\varepsilon4$ none; Vitamin B12 deficiency absent.\\
Prior sets:\\
\item  \textbf{(A) Baseline}: $b \sim \mathcal{N}(0,1)$; Intercept $\sim \mathcal{N}(0,2)$; $\mathrm{SD}_{\mathrm{ADC}} \sim \mathrm{Exp}(1)$; Weibull shape $\sim \mathrm{Exp}(1)$; log-logistic $\sigma \sim \mathrm{Exp}(1)$.
\item \textbf{(B) Heavy-tail}: $\mathrm{SD}_{\mathrm{ADC}} \sim \mathrm{Exp}(0.5)$; Weibull shape $\sim \mathrm{Exp}(0.5)$; log-logistic $\sigma \sim \mathrm{Exp}(0.5)$.
\item \textbf{(C) More shrinkage}: $\mathrm{SD}_{\mathrm{ADC}} \sim \mathrm{Exp}(2)$; log-logistic $\sigma \sim \mathrm{Exp}(2)$.
\item \textbf{(D) Half-normal RE}: $\mathrm{SD}_{\mathrm{ADC}} \sim \mathcal{N}^{+}(0,1)$; Weibull shape $\sim \mathrm{LogNormal}(0,0.5)$; log-logistic $\sigma \sim \mathcal{N}^{+}(0,1)$.
\item \textbf{(E) Wide intercept}: Intercept $\sim \mathcal{N}(0,5)$; $\mathrm{SD}_{\mathrm{ADC}} \sim \mathcal{N}^{+}(0,2)$; Weibull shape $\sim \mathrm{LogNormal}(0,1)$; log-logistic $\sigma \sim \mathcal{N}^{+}(0,2)$.
\item \textbf{(F) Strong slopes}: $b \sim \mathcal{N}(0,0.5)$; Weibull shape $\sim \mathrm{LogNormal}(0,0.5)$; log-logistic $\sigma \sim \mathrm{LogNormal}(0,0.5)$.
\item \textbf{(G) Student-$t$ slopes}: $b \sim t_{3}(0,1)$; Intercept $\sim \mathcal{N}(0,5)$; $\mathrm{SD}_{\mathrm{ADC}} \sim \mathcal{N}^{+}(0,1)$; Weibull shape $\sim \mathrm{LogNormal}(0,0.5)$; log-logistic $\sigma \sim \mathrm{LogNormal}(0,0.5)$.
\end{tablenotes}

\end{threeparttable}
\end{table*}

\end{landscape}
\clearpage

\begin{figure}[!ht]
\centering
\begin{subfigure}[t]{0.48\textwidth}
  \centering
  \includegraphics[width=\textwidth]{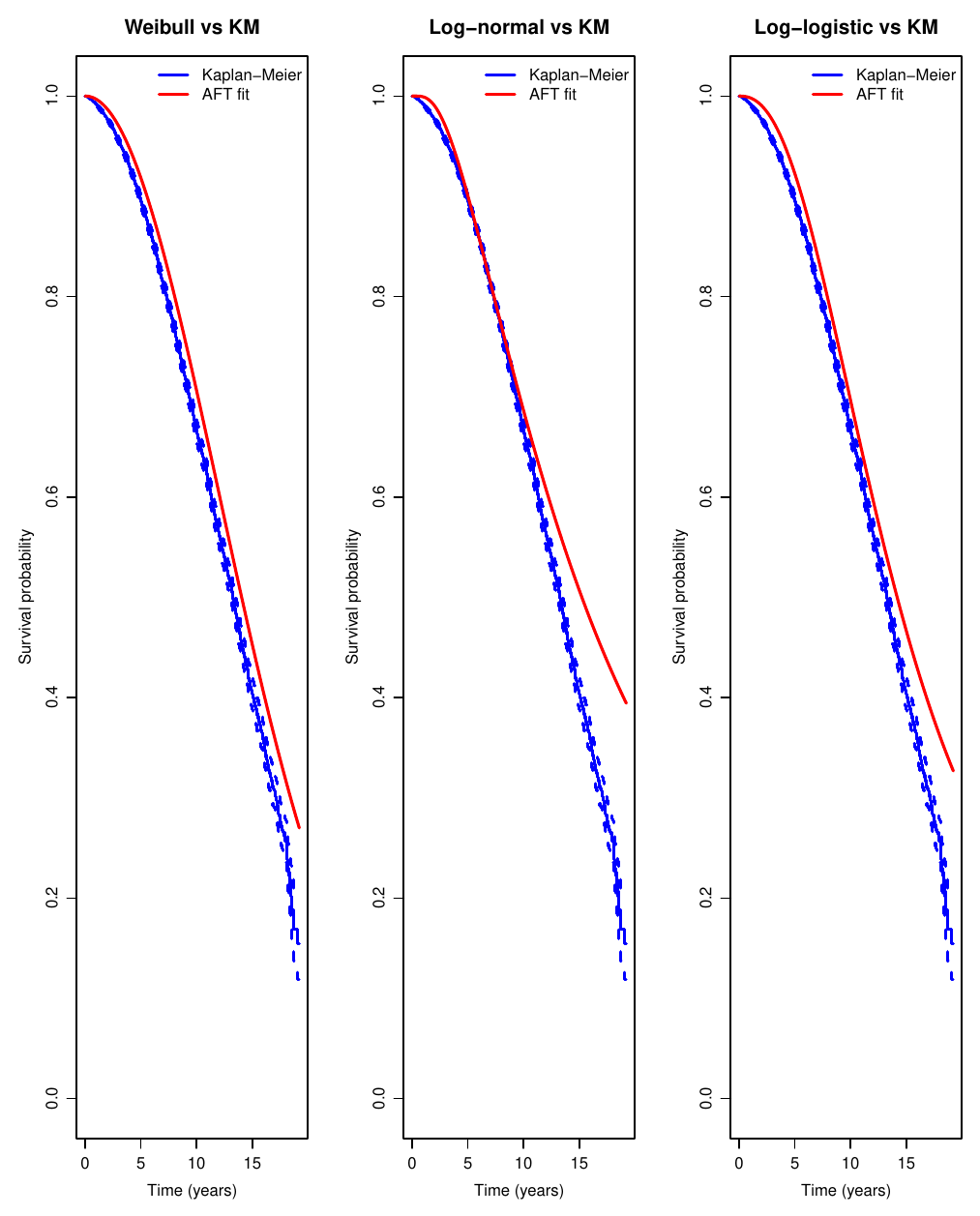}
  \caption{Time to first MRI (Cause 1).}
  \label{fig:aft-mri}
\end{subfigure}
\hfill
\begin{subfigure}[t]{0.48\textwidth}
  \centering
  \includegraphics[width=\textwidth]{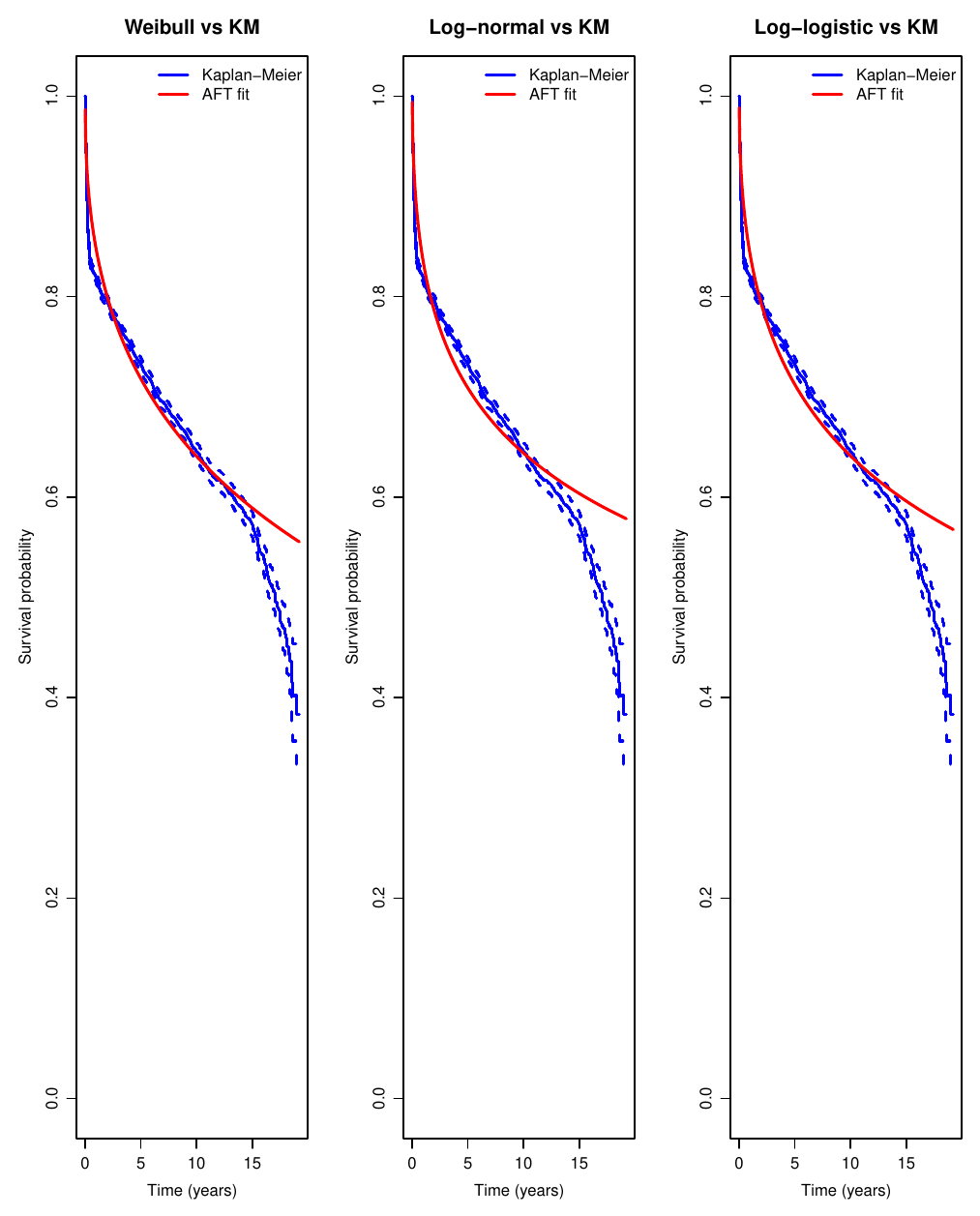}
  \caption{Time to death before MRI (Cause 2).}
  \label{fig:aft-death}
\end{subfigure}
\caption{Model diagnostics for competing-risk AFT models. (A) Comparison of Kaplan-Meier survival with fitted Weibull, log-normal, and log-logistic models for time to first MRI (Cause 1); the Weibull fit provided the best overall agreement and lowest AIC/BIC. (B) Corresponding diagnostics for time to death before MRI (Cause 2); the log-logistic distribution captured the heavy tail of mortality and minimized AIC/BIC.}
\label{fig:aft-both_var}
\end{figure}

\begin{figure}[!ht]
\centering
\includegraphics[width=0.8\textwidth]{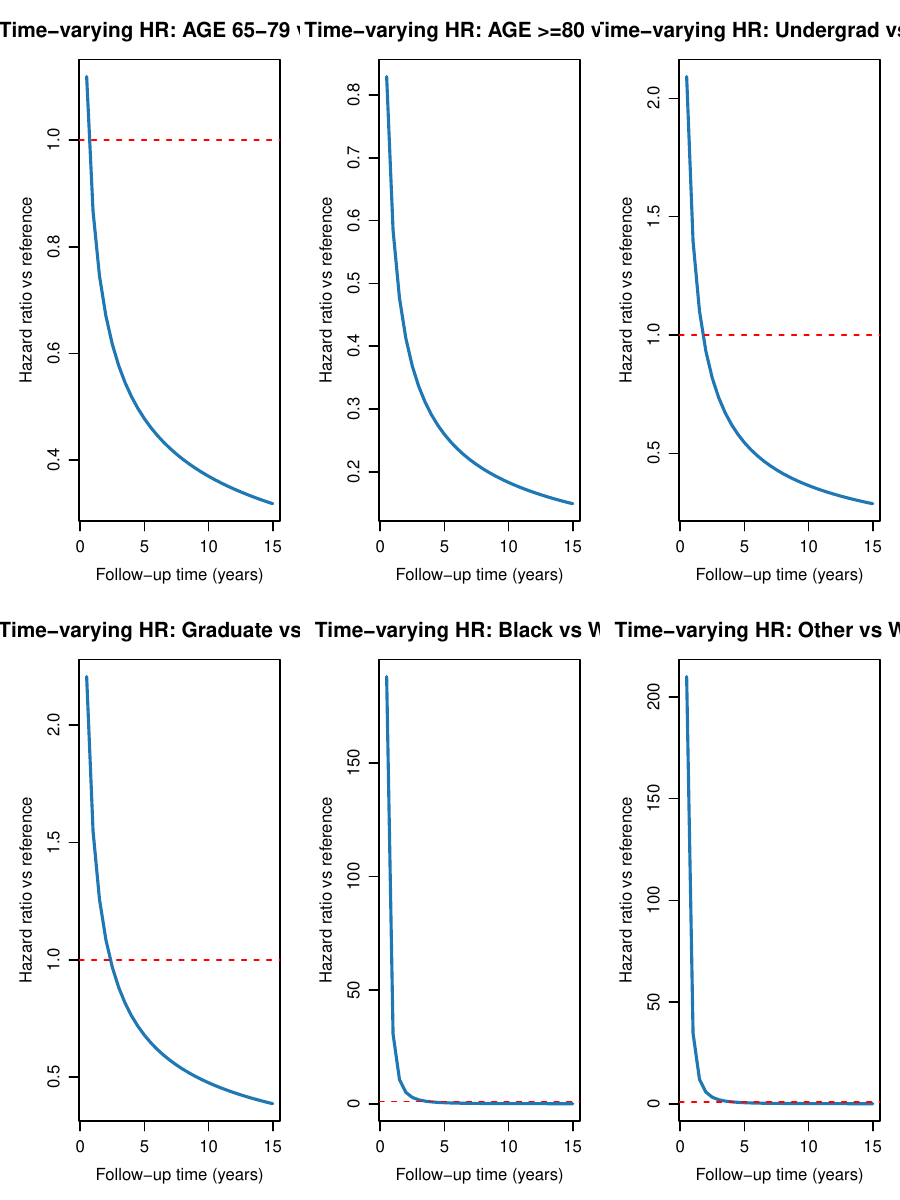}
\caption{Time-varying hazard ratios from the cause--specific TVC Cox model for first MRI: age groups, education groups, and race/ethnicity. Dashed red lines indicate HR = 1 (no difference vs.\ reference).}
\label{fig:tvc}
\end{figure}

\begin{figure}[!ht]
\centering
\begin{subfigure}[t]{0.48\textwidth}
  \centering
  \includegraphics[width=\textwidth]{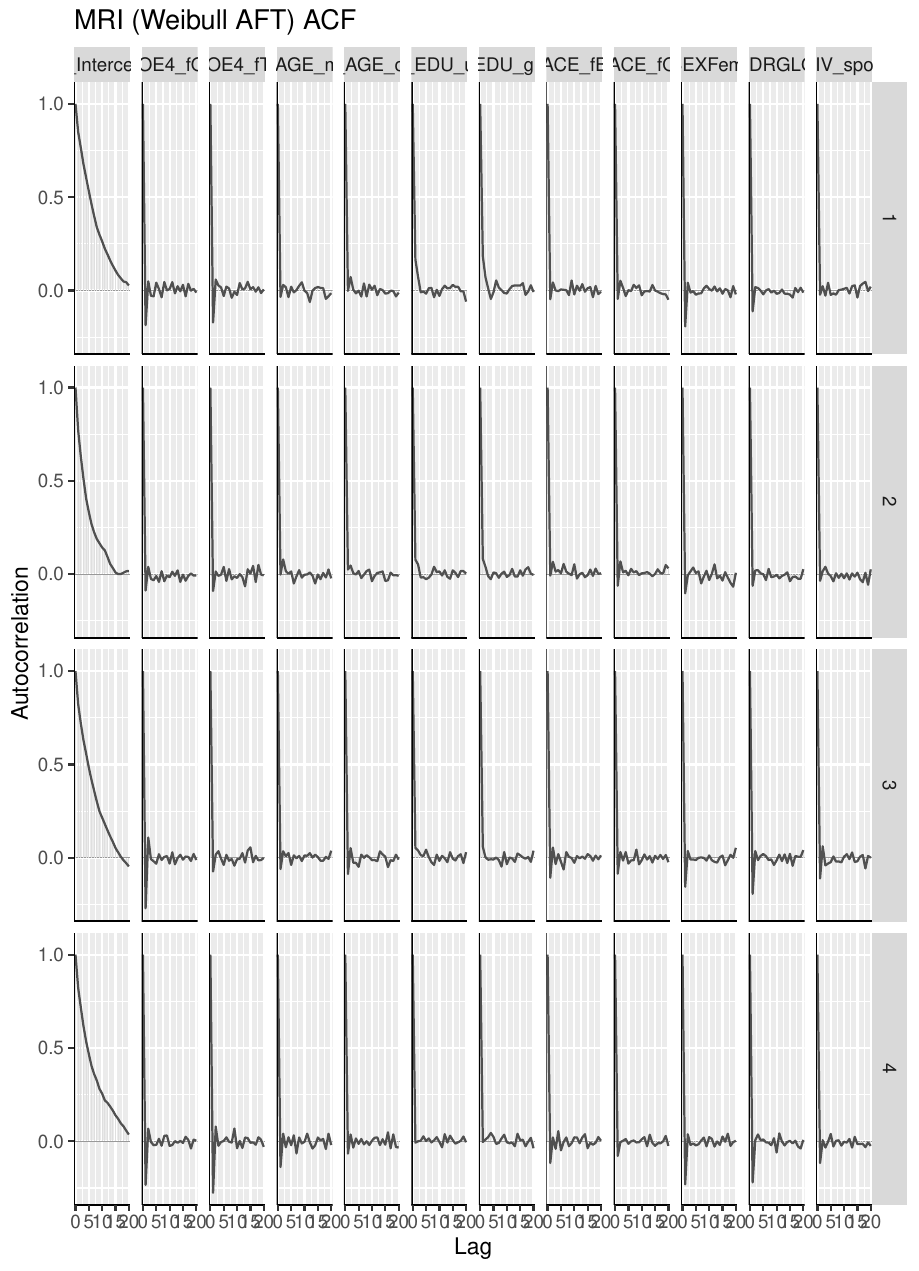}
  \caption{Time to first MRI (Cause 1).}
  \label{fig:aft-mri_acf}
\end{subfigure}
\hfill
\begin{subfigure}[t]{0.48\textwidth}
  \centering
  \includegraphics[width=\textwidth]{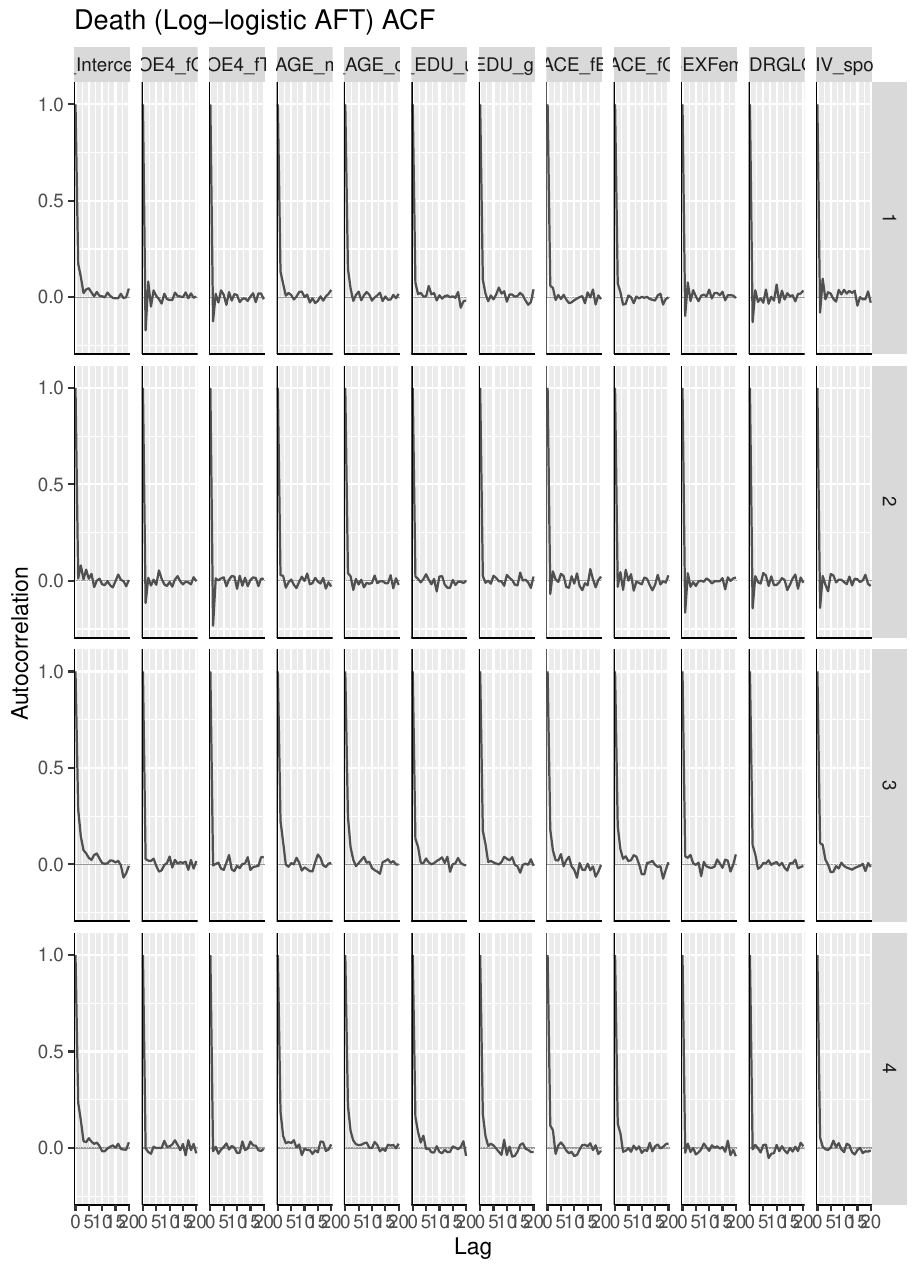}
  \caption{Time to death before MRI (Cause 2).}
  \label{fig:aft-death_acf}
\end{subfigure}
\caption{Model diagnostics for competing-risk AFT models. (A) Comparison of Kaplan-Meier survival with fitted Weibull, log-normal, and log-logistic models for time to first MRI (Cause 1); the Weibull fit provided the best overall agreement and lowest AIC/BIC. (B) Corresponding diagnostics for time to death before MRI (Cause 2); the log-logistic distribution captured the heavy tail of mortality and minimized AIC/BIC.}
\label{fig:aft-both_fg}
\end{figure}

\begin{figure}[!ht]
\centering
\begin{subfigure}[t]{0.485\textwidth}
  \centering
  \includegraphics[width=\textwidth]{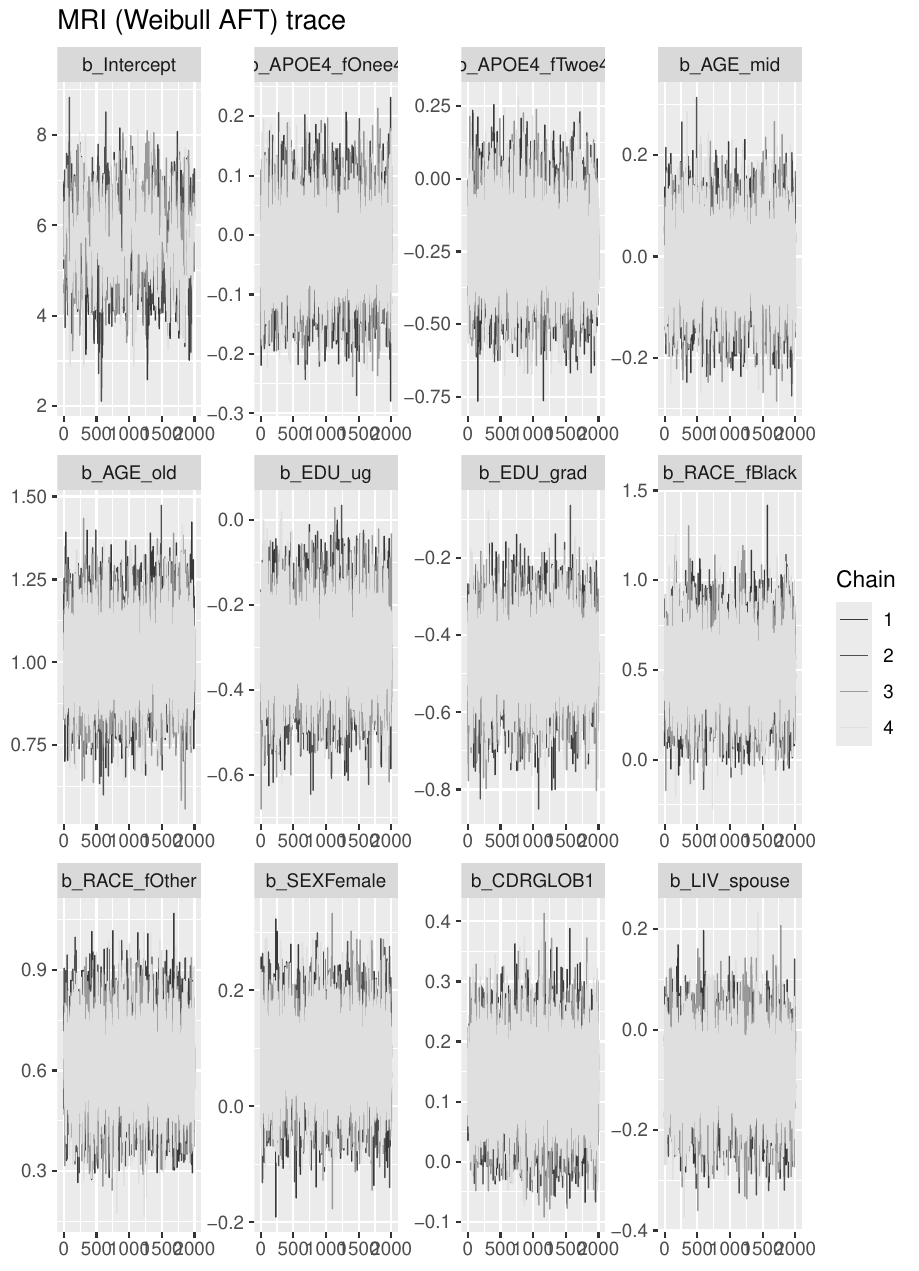}
  \caption{Time to first MRI (Cause 1).}
  \label{fig:aft-mri_trace}
\end{subfigure}
\hfill
\begin{subfigure}[t]{0.485\textwidth}
  \centering
  \includegraphics[width=\textwidth]{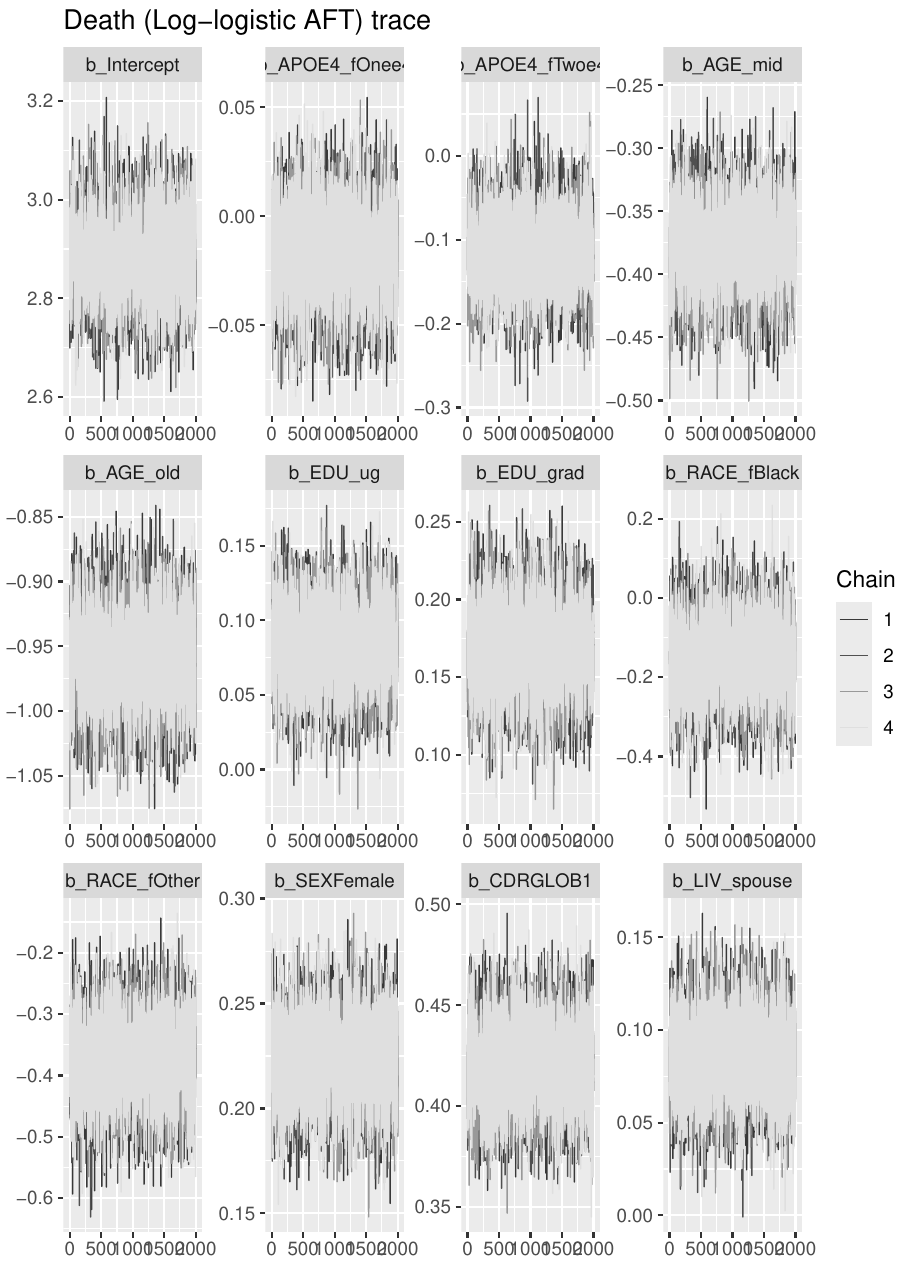}
  \caption{Time to death before MRI (Cause 2).}
  \label{fig:aft-death_trace}
\end{subfigure}
\caption{MCMC trace plots for competing-risk AFT models. (A) Trace plots for the time-to-first-MRI (Cause 1) Weibull model; all four chains mix well and remain stationary across parameters, indicating good convergence. (B) Trace plots for the time-to-death-before-MRI (Cause 2) log-logistic model, showing similarly stable mixing across chains, supporting reliable posterior estimates for both cause-specific models.}
\label{fig:aft-both_tace}
\end{figure}

\begin{figure}[!ht]
\centering
\begin{subfigure}[t]{0.485\textwidth}
  \centering
  \includegraphics[width=\textwidth]{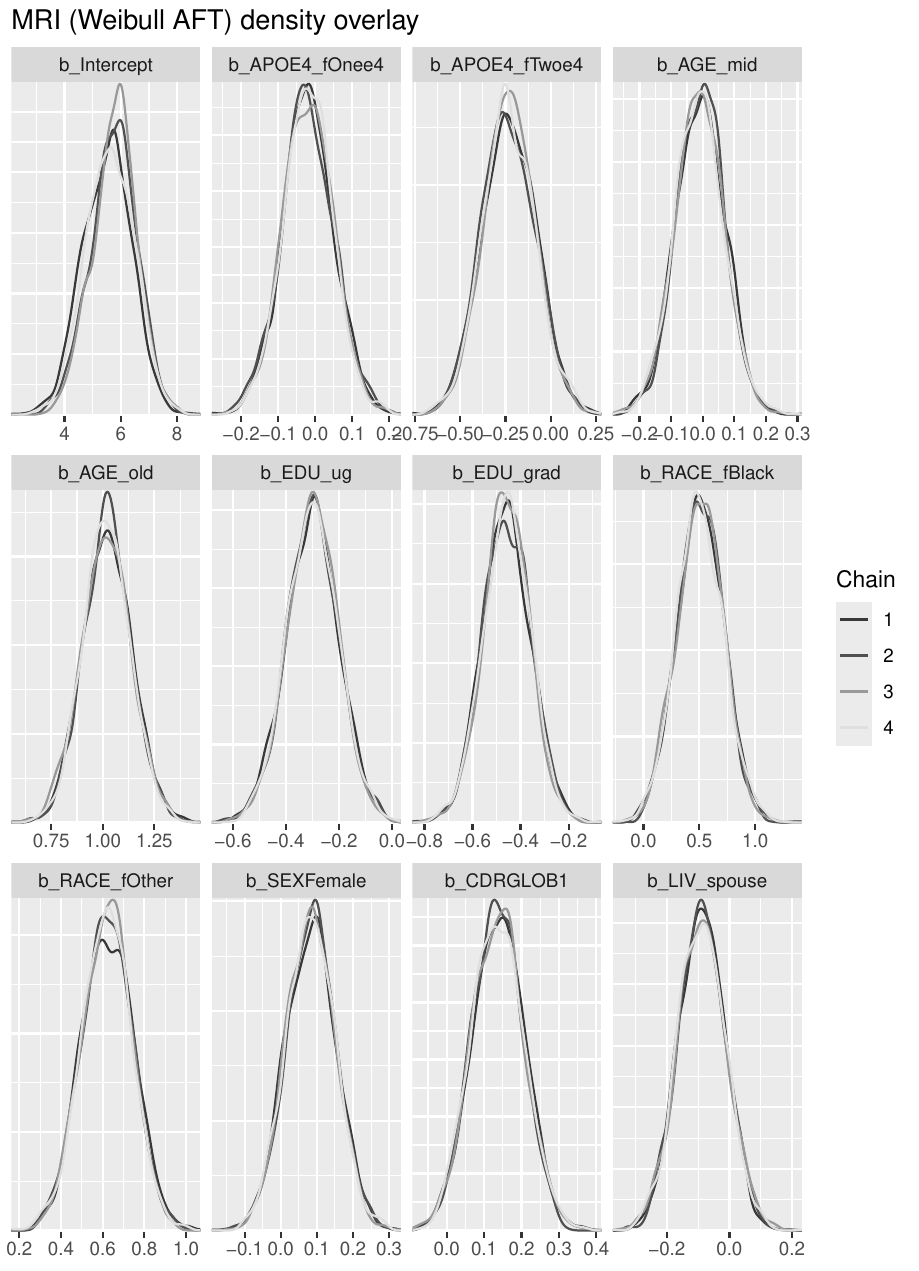}
  \caption{Time to first MRI (Cause 1).}
  \label{fig:aft-mri_over}
\end{subfigure}
\hfill
\begin{subfigure}[t]{0.485\textwidth}
  \centering
  \includegraphics[width=\textwidth]{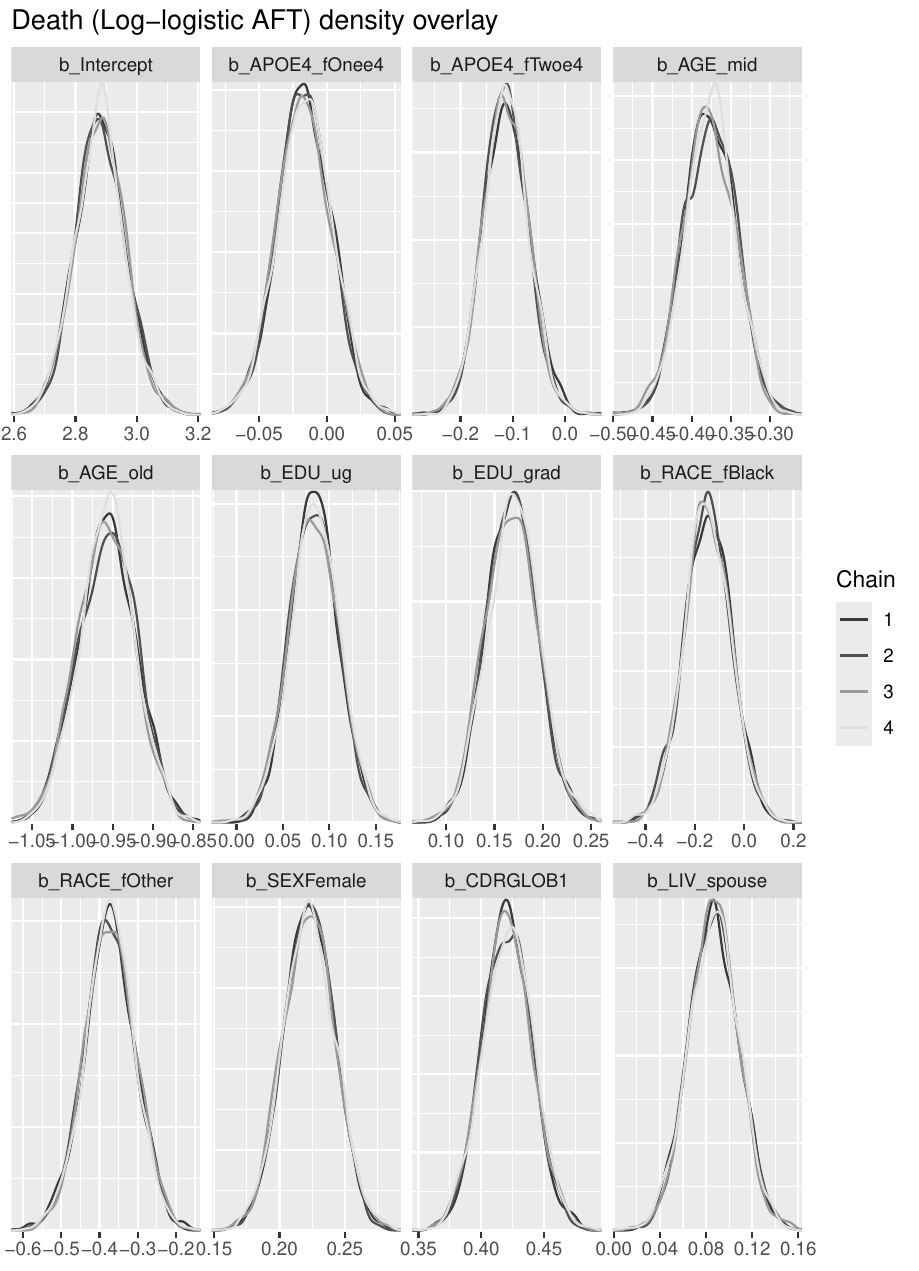}
  \caption{Time to death before MRI (Cause 2).}
  \label{fig:aft-death_over}
\end{subfigure}
\caption{Posterior density overlays for competing-risk AFT models. (A) Density plots for the time-to-first-MRI (Cause 1) Weibull model; chains overlap closely with unimodal distributions, indicating good convergence. (B) Density plots for the time-to-death-before-MRI (Cause 2) log-logistic model, showing similarly consistent agreement across chains.}
\label{fig:aft-both_all}
\end{figure}

\end{document}